%Paper: astro-ph/9403028
%From: <fbernard@chipmunk.cita.utoronto.ca>
%Date: Fri, 11 Mar 94 20:15:03 EST

%                                                       March 1994
% `Properties of the Cosmological Density PDF' by F. Bernardeau and L. Kofman
% Plain Tex format.
% table and figures available by anonymous ftp at
%                    ftp.cita.utoronto.ca
% in the directory
%                    cita/francis/lev
%
%\magnification\magstep1
\headline={\ifnum\pageno=1\hfil\else\hfil\tenrm--\ \folio\ --\hfil\fi}
\footline={\hfil}
\baselineskip 12pt
\parskip=6pt
%\voffset -1.5truecm
%Comment this line out if using postscript printer; uncomment it for imagen.
%\hsize=6.4 truein \vsize=8.9 truein \hoffset=1 truein \voffset=1 truein

\tolerance=10000 \hyphenpenalty10000 \exhyphenpenalty10000
\def\pp{\parshape 2 0truecm 15truecm 2truecm 13truecm}
\def\apjref#1;#2;#3;#4; {\par\pp#1, {#2}, {#3}, #4}
\def\bookref#1;#2;#3; {\par\pp#1, { #2}, {\rm #3}}
\def\prepref#1;#2; {\par\pp#1, {#2}}
\overfullrule=0pt
\def\cl{\centerline}
\def\page{\vfill\eject}

\def\etal{{et al.\ }}

\def\grad{\nabla}
\def\ltsima{$\; \buildrel < \over \sim \;$}
\def\lsim{\lower.5ex\hbox{\ltsima}}
\def\gtsima{$\; \buildrel > \over \sim \;$}
\def\gsim{\lower.5ex\hbox{\gtsima}}
\def\vv{\hbox{\bf v}}
\def\vu{\hbox{\bf u}}
\def\vx{\hbox{\bf x}}

\def\vq{\hbox{\bf q}}
\def\vS{\hbox{\bf S}}
\def\d{\hbox{d}}
\def\i{\hbox{i}}
\def\mM{{\cal M}}
\def\mC{{\cal C}}
\def\mG{{\cal G}}

\def\mg{\big<}
\def\md{\big>}
\def\dta{\delta}
\def\rhos{\varrho}
\def\lam{\lambda}

\vbox{\vskip 3truecm}
\bigskip
\cl{\bf PROPERTIES  OF THE COSMOLOGICAL DENSITY}
\cl{\bf DISTRIBUTION FUNCTION}
\vskip 1.0 truecm
\bigskip\bigskip\bigskip
\cl{by}
\bigskip
\cl{\bf Francis Bernardeau}
\bigskip
\cl{ CITA, University of Toronto, Toronto, ON M5S 1A7,Canada}
\bigskip\bigskip
\cl{and}
\bigskip\bigskip
\cl{\bf Lev Kofman}
\bigskip
\cl{Institute for Astronomy, University of Hawaii, Honolulu, HI 96822}
\bigskip\bigskip\bigskip\bigskip\bigskip
\cl{  Submitted to: {\it  The Astrophysical Journal}}

\page
\vglue 3 truecm
\cl{\bf ABSTRACT}

The  properties of the probability distribution function of the cosmological
continuous density field are studied. We present further developments and
compare dynamically motivated methods to derive the PDF. One of them is based
on the Zel'dovich approximation (ZA). We extend this method for arbitrary
initial conditions, regardless of whether they are  Gaussian or not. The other
approach is based on perturbation theory with Gaussian initial fluctuations.
We include the smoothing effects in the PDFs.

 We examine the relationships
between the shapes of the PDFs and the moments. It is found that formally there
are no moments in the ZA, but a way  to resolve this issue is proposed, based
on the regularization of integrals. A closed form for the generating function
of the moments in the ZA is also presented, including  the smoothing effects.
We suggest the methods to build PDFs out of the whole series of the moments,
or out of a limited  number of moments -- the Edgeworth expansion. The last
approach gives us an alternative method to evaluate the skewness and kurtosis
by measuring the PDF around its peak. We note a general connection between the
generating function of moments for small r.m.s $\sigma$ and the non-linear
evolution of the overdense spherical fluctuation in the dynamical models.

 All these approaches have been applied
in 1D case where the ZA is exact, and simple
analytical results are obtained. It allows us to study in details how these
methods are related to each other. The 3D case is analyzed in  the same manner
and  we found a mutual agreement in the PDFs derived by different methods in
the the quasi-linear regime.  Numerical CDM simulation was used to validate the
accuracy of considered approximations. We explain  the successful  log-normal
fit of the PDF from that simulation at moderate $\sigma$ as mere fortune, but
not as a universal form of density PDF in general.

{\it Subject headings:}\ cosmology: theory --- dark matter --- galaxies:
clustering

\page

\centerline{\bf 1. INTRODUCTION}
\medskip

% statistics

One of the  goals of the study of the nonlinear gravitational
 dynamics in an expanding universe is the determination
of the statistical properties of the various cosmic fields that can be used
to describe the matter distribution and motion.
 In the linear regime, for Gaussian initial
conditions, this description is quite easy since it reduces to the
behavior of the two-point correlation function, or equivalently to
the shape of the power spectrum.
Once nonlinear effects are taken into account this is no longer the case
and the mathematical description of the statistical properties
is more complicated.
One can basically distinguish two approaches to describe the
statistics of the non-linearities.

 The description of statistics in terms
of the correlation functions
 has been investigated  since '70s. In principle, a full knowledge of the
matter field can be obtained from the shape of the $p$-point correlation
functions (see Peebles 1980 for references).
These functions are the solution of a hierarchy of dynamical  equations,
 the
BBGKY hierarchy.
The  spatial correlation functions  have been measured in
galaxy catalogues, or in numerical simulations.
Progress was made  to find a few low order correlation functions,
 theoretically, numerically and observationally
  (see Peebles 1993 for references).
However, dynamical BBGKY equations have never been solved in general.

Another approach which drew much attention   recently is based on the
 probability distribution function (PDF) of a cosmic field at a given point
 or  PDFs at various points, or joint PDFs of several cosmic fields.
 The one-point density PDF,
$P(\rho)\,\d\rho,$
 is the primary object of study in this approach.
The discrete analogy of the one point PDF is given by counts in cell
probabilities that is basically what is obtained after
smoothing of the discrete field by a sharp filter.
It is important to note that
the PDFs contain more statistical information than a few lower order
correlation functions. Actually the moments of the PDF are the spatial
averages (with the same window functions) of the correlation functions.

In practice,
galaxy PDFs have been measured in various catalogues by
Hamilton (1985) and
Alimi, Blanchard \& Schaeffer (1990) in the CfA survey, Bouchet \etal (1993)
Kofman et al. (1994)  in the IRAS surveys,
Maurogordato, Schaeffer \& Da Costa (1992) in the
SSRS survey, Gazta\~naga \& Yokoyama (1993) both in the SSRS and
CfA survey.  The density PDF manifests significant non-Gaussian
features in non-linear and even in quasi-linear regimes.
One central question theories have to address is to determine
what part of this non-Gaussianity came from
non-linear dynamics,  what part  came from possible  non-Gaussian initial
 conditions, and what part came from the galaxy biasing.

To see what kind of density PDFs  emerge from gravitational dynamics,
 investigations have been made
in numerical simulations by
Bouchet, Schaeffer \& Davis (1991), Weinberg and Cole (1992),
 Bouchet \& Hernquist (1993),
Juszkiewicz \etal (1993), Kofman et al. (1994).

In this paper we concentrate on the theoretical basis for the
derivation of the cosmic density PDF.
There were phenomenological attempts in the literature {\it  to design}
 $P(\rho)$ -- see, for instance, Saslaw and Hamilton (1984); Coles and Jones
(1991).
Here we discuss {\it  the derivation} of the density PDF from gravitational
dynamics, and its comparison against  numerical simulation.
It is a difficult problem to derive $P(\rho)$ for the general case.
However, it can be studied in different regimes and approximations.
One can distinguish different successive  stages of the non-linear
 gravitational dynamics:  quasi-linear regime, non-linear regime,
 and highly non-linear regime.
The quasi-linear (or mildly non-linear) regime when $\sigma < 1$ can be
 investigated by the mean of
the perturbation theory, unlike other regimes.
In the non-linear regime $\sigma > 1$ the complexity of the dynamics makes
 analytical studies virtually impossible.
 Highly non-linear regime when  $ \sigma \gg 1$ might take place
when the  hypothetic hierarchical ansatz for the correlation functions
 is working. Indeed, another feature
of the gravitational dynamics -- self-similar solution
(Davis \& Peebles 1977)-- is
expected to be reached in this  regime.
The statistics of this regime has
been a subject of consideration in many papers.
In particular Balian \& Schaeffer (1989) give
relations of crucial interest that relate the properties of
one-point density PDF to the ones of the correlation functions.

In this paper, we rather concentrate on
the statistics in the quasi-linear  regime.
Fortunately, a fair fraction of the observational data does correspond to such
a regime when the galaxy surveys are smoothed with a large enough
radius (say $\gsim 8 h^{-1}$Mpc).
In the quasi-linear regime the two--point correlation function is mainly
 determined
by the initial conditions. The manifestation of the nonlinear features
will be seen in the higher--order correlation functions,
or equivalently in the departure of the shape of the distribution
functions from Gaussian distributions.
We have now various
tools at our disposal to study the mildly
nonlinear regime  in details, so that specific predictions can be made
that are directly derived from first principles.

One of the most successful approximations to describe the early
 nonlinear dynamics
is the Zel'dovich approximation (Zel'dovich 1970).
In this approximation, the displacement of
the particles is extrapolated from its behavior in the linear regime.
The whole local statistical properties of the
field at the position of a particle is then determined by the
statistical properties of the initial displacement field.
For Gaussian initial condition, it is possible, for instance, to compute
the density PDF by a change of variable starting with the initial
statistical properties of the displacement. Actually, as it will be recalled
in this paper,
a lot more statistical properties can be derived with such
an approximation (Kofman 1991, Kofman \etal 1994).

The second method is based on the calculations of the large--scale
cumulants of the cosmic field from perturbation theory.
Indeed, when one assumes Gaussian
initial conditions, it turns out that it is possible to derive the
leading term (with respect to the rms density
fluctuation) of each cumulant (Bernardeau 1992).
These first order contributions are expected
to dominate the exact value of the cumulants as long as the rms
fluctuation is accurately given by the linear theory.
The shape of the density PDF is then obtained
by a reconstruction of this function from the generating function
of the moments.

The mathematical forms of the density PDFs derived analytically
 in these two methods are quite
different. We compare the forms of the PDFs derived in the quasi-linear regime
 earlier
in our papers, and show its mutual agreement.
 As this subject is in rapid development,
many questions and controversies
on the density PDF and moments are accumulating in the literature.
 Therefore we try to clarify many points
related to the PDFs, generating functions, moments and smoothing in
the quasi-linear regime.
 Presumably, there is no simple
 universal formula  for $P(\rho)$ in general.
However, one of the practical outputs
of our paper is to provide a justification for the
 use a log-normal distribution as
a successful fit for the actual cosmic density PDF.
This fit is only accurate for
the cosmological  models based on the Gaussian initial statistics
and realistic power spectrum, but not in the general case.
Another practical output is a new view on the measurement of the lowest
 moments - skewness and kurtosis. Common belief is that they are affected
by the high density tail which is difficult to measure. We will
demonstrate that these moments can be evaluated from the form of the  PDF
around its maximum.

We will use the concepts of moments, cumulants, PDFs and generating function
throughout the paper. The formal definitions of all of them are collected
 in
Appendix A. However, we must  stress that the introduction of  quantities
such
as the generating functions of moments and cumulants
is not an extra mathematical exercise but
an useful tool to find the measurable statistics of the cosmic fields.
There is a deep connection between generating function and the non-linear
evolution
of the spherical overdense fluctuation in the dynamical model of the
gravitational instability.
Hence, in most cases the generating function
can be derived directly from the dynamical equations.
 We will present the closed forms of the generating functions
in different approximations in the quasi-linear regime.

Throughout this analysis we got a series of new results:
the generating functions of cumulants and finite moments in the Zel'dovich
approximation, both with and without final smoothing;
two methods of reconstruction of cosmological density PDF from,
 correspondingly, the whole (Laplace transformation) and partial (Edgeworth
decomposition) series of moments; systematic mutual comparison of  all these
methods and their comparison against the calculations from N-body simulation;
the range of validity of their fitting by the log-normal distribution;
the extension of the derivation of density PDF in the Zel'dovich approximation
for an arbitrary initial condition.

Section 2 of the paper is devoted to the 1D dynamics, in which the mathematical
content is simpler than for the 3D dynamics. We apply  general methods
and mathematical tools in this case.
Section 3 is devoted to the statistical properties that can be derived
from the use of the Zel'dovich approximation for  the 3D dynamics.
In Section 4, we give the properties of the exact dynamics derived
from direct perturbative calculation in the single stream regime.
 Section 5 is devoted to comparisons
with numerical simulations. In the conclusive Section 6 we summarize the
approximations that have been made and the results in a detailed table.

\vskip 1 cm
\centerline{\bf 2. AN ILLUSTRATIVE EXAMPLE: THE 1D DYNAMICS}
\vskip .2 cm

In this part we aim to present the analytical tools developed
in the mildly nonlinear regime in case of a simpler dynamics.
We assume in this part  that the fluctuations are only one-dimensional.
The general methods suggested here will be used in the
 next sections in the 3D case.

\vskip .5 cm
\leftline{\sl 2.1. The construction of the density PDF from the
Zel'dovich approximation}

It is convenient to use the  equation of motion  in the  Lagrangian
 description.
In such a case the dynamics is described by the displacement, $\vS(\vq, t)$,
of each particle from its initial position $\vq$. Its current Eulerian
comoving position,
$\vx$, is then given by
$$\vx=\vq+\vS(\vq, t).\eqno(1)$$
For 1D case the mildly non-linear dynamics is quite simple. The reason is
 that the force
exerted by a density perturbation over a given particle
is independent of its distance to  the particle.
 Therefore,
before any shell crossing, the displacement of  each particle depends on
its Lagrangian position $\vq$ only. Then
 the displacement field for the growing mode can be factorized
$$\vS(\vq,t)=\Psi(\vq)\ D(t).\eqno(2)$$
The 3D generalization of this form of the displacement is the Zel'dovich
approximation (ZA). In 1D case the ZA is then identical to the exact dynamics.
In relation (2) the displacement factorizes in a spatial function,
$\Psi(\vq)$, which depends on the  initial conditions, and a universal
time dependent function $D(t)$ which contains by definition the time
dependence of the growing modes.
Let us call $\lambda_0$ the  derivative of $-\Psi(\vq)$ with respect
to  $\vq$.
The local comoving
density (in units of the mean density,
$\varrho={\rho / \bar \rho}$)  is then, by the mass conservation
 equation, given by
$$\varrho(\vq)={1\over\vert 1-D\lambda_0\vert}.\eqno(3)$$

The reconstruction of the density PDF then is based on the statistical
properties of $\lambda_0$. For Gaussian initial conditions, $\lambda_0$
simply obeys  Gaussian statistics. The derivation of the shape
of the density PDF is then just a matter of  change
of variable, and as  it can be seen in (3) the density contrast
will not be normally distributed. This is  general feature due to the
nature of this quantity. The positive density contrast can reach any
large value while the negative density contrast is restricted by
 $\varrho \ge 0$.
Hence the probability function $P(\varrho, D(t))$, meaning the fraction of
volume with a given value of density, is expected to be very non-Gaussian
even in the quasilinear stage. However, note that a simple change of
variable leads to the density PDF at a given point, in Lagrangian space.
The distribution in Eulerian space should take into account the fact that
a given amount of matter in a dense spot occupies a small volume.
The density PDF in Eulerian space is then obtained by divided the
density distribution in Lagrangian space by the density, and multiplied
by the numerical factor  which controls the normalization.
This factor is related to the number of streams $\mg N_s \md$,
 in cases $\sigma \lsim 1$
under consideration it is very close to unity, and we ignore it
(e.g. Gurbatov et al. 1991).

Then the density PDF in 1D reads
$$\eqalign{
P_{1D}(\varrho)\d\varrho&={1\over (2 \pi\sigma^2)^{1/2}}
\left[\exp(-{\lambda^2\over2\sigma^2})
+\exp(-{\lambda'^2\over2\sigma^2})\right]{\d\varrho\over\varrho^3},\cr
\lambda \equiv D \,\lambda_0&=1-{1\over\varrho},\ \ \ \lambda'
 \equiv D \,\lambda_0'
=1+{1\over\varrho}.\cr}
\eqno(4)$$
Note that $\sigma=\sigma_0 \, D(t)$, where
$\sigma_0$ is the variance of the initial linear density contrast.
In the limit of small $\sigma$ the distribution (4) transits to the
Gaussian form.  The high density asymptota $\varrho^{-3}$ that is seen
in equation (4) is induced by the caustics.

\vskip .5 cm
\leftline{\sl 2.2. Density PDF and moments}

As recalled in the introduction, the knowledge of the shape of the
PDF and the moments are intimately related. The very example we are presenting
now is here to point out that this is not completely true. Indeed, none
of the moments of the previous distribution are finite,
 because of the  $\varrho^{-3}$-asymptota!
 It can be easily checked that even for an arbitrary small $\sigma$
  the density PDF behaves like $\sqrt{2 /\pi}/\sigma
\exp(-1/2\sigma^2) \varrho^{-3}$ at high density.
However, physical density in caustics is finite.
 We expect that a physical process will operate as sort of regularization
of the high density tail of $P(\varrho)$.
For example, Zel'dovich \&  Shandarin (1982) showed  how physical
properties of the hot dark matter lead to the finite density in the caustics.
 Obviously we do not know what will be the exact form of this regularization
in general case,
so we simply describe the regularization process by a cut-off at large
density (where the physical regularizing processes are thought to
occur) that makes the moments finite.
The shape of this cut-off is somewhat arbitrary, but we will see that,
to some extent,
a fair fraction of the properties of the moments do not depend on the
procedure that has been adopted.
We chose two  toy models of distribution,
with the sharp cutoff and with the exponential cutoff, in the following form:
$$\eqalign{
P_{\rm reg1}(\varrho)\ \d\varrho&
={1\over C_1}P_{1D}(\varrho)\ \d\varrho\ \ \ \ \hbox{if}
\ \ \ \ \varrho<\varrho_c,\cr
P_{\rm reg1}(\varrho)\ \d\varrho&=0\ \ \ \ \hbox{if}
\ \ \ \ \varrho>\varrho_c,\cr}\eqno(5)$$
and
$$\eqalign{
P_{\rm reg2}(\varrho)\ \d\varrho&={1\over C_2}\
P_{1D}(\varrho)\left[1+{\exp(x)\over\exp(x)+\exp(-x)}\left(
\exp(-\varrho/\varrho_c)-1\right)\right]\ \d\varrho,\cr
x&=\varrho-\varrho_c.\cr}\eqno(6)$$
The coefficients $C_1$ and $C_2$ are simply normalization factors.
The adopted values of $\varrho_c$ will be quite large (above 5)
insuring that the shape of the density PDF is not changed in the domain
of interest that is for $0<\varrho< $ several.
The coefficients $C_1$ and $C_2$ are very close to unity in these cases.
In Fig. 1 we present the distribution (4) and regularized
distributions (6).

Once such a regularization is made, the moments of the distribution functions
${\mg\delta^p\md}=\int_0^{\infty}\d\varrho\ P(\varrho)(\varrho-1)^p$
can be easily calculated numerically.
We  present the results in terms of $S_p$ parameters adopted in the
literature. These  parameters are defined through the cumulants
(see Appendix A, Eq. [A2]):
$$S_p=\mg\dta^p\md_c/\sigma^{2(p-1)}.\eqno(7)$$
This is a generic definition which is not based on any a priori assumption.
The $S_p$ parameters would be constants in the hierarchical ansatz,
but in general they can be seen as functions of $\sigma$.

The parameter $S_3$ multiplied by $\sigma$ is the skewness,
and $S_4$ multiplied by $\sigma^2$ is the kurtosis.
The numerical calculations of the parameters $S_3$ and $S_4$ for
the regularized    distributions (5) and (6) for the exact Zel'dovich solution
is shown in Fig. 2.
The measurements have been made by calculating the actual second, third,
and fourth moment of the regularized density PDF, and by taking the
appropriate ratios.
We see that these coefficients are finite. The shapes of  $S_3(\sigma)$ and
 $S_4(\sigma)$ are universal in the interval of $\sigma$ from 0 up
to the value $\sim 0.2$. These shapes are independent  of the form
 of the cutoff, as well as on the parameter of truncation.
In the limit of $\sigma \to 0$ we found $S_3 \approx 6$, and $S_4 \approx 72$.
In the next section we will confirm these figures and their universality
 by analytical calculations.
We conclude that the found values of  $S_3(\sigma)$ and $S_4(\sigma)$
 for $\sigma  \lsim 0.2$ are the proper moments in the 1D case.
There is also some universality of these coefficients for $\sigma \gsim 1.2$,
in the sense that it does not depend on the parameter $\varrho_c$
for each panel.
However, it does depend on the form of regularization. We suggest that
the found values of   $S_3(\sigma)$ and  $S_4(\sigma)$ for $\sigma \gsim 1.2$
do not reflect the proper moments, but rather the shape of the adopted cutoff
and thus do not have any physical meaning.

%\page
\vskip .5 cm
\leftline{\sl 2.3. Calculation of  $ S_p$ for small $\sigma$}

It is possible
to calculate the $S_p$ series analytically when the variance $\sigma$ of the
distribution function is small.
This calculation requires the introduction of the generating
functions of moments and cumulants (see Appendix A).
The formal definition of  the generating function of the moments
$ \mg\dta^p\md$ is
$$\mM(\mu)=1+\sum_{p=1}^{\infty}\mg\dta^p\md {\mu^p
\over p!}\eqno(8)$$
and the generating function of the cumulants $\mg\dta^p\md_c$ is
$$\mC(\mu)=\sum_{p=2}^{\infty}\mg\dta^p\md_c
{\mu^p\over p!}.\eqno(9)$$
Here $\mu$ is an auxiliary parameter.
Using its definition, one can relate $\mM(\mu)$ to the shape of the
density PDF:
$$\mM(\mu)=\exp{\mC(\mu)}=
\int_0^{\infty}\d\varrho\ P(\varrho)\exp([\varrho-1]\mu).\eqno(10)$$

Since we know the density PDF  in the quasi-linear regime in the 1D problem,
in principle,  we can obtain the generating function $\mC(\mu)$
 substituting the formula (4) into integral (10).
When $\sigma \ll 1$  we can integrate (10) in a
 closed form, using the steepest descent method.
The saddle point of the exponent  is given by the equation
$$-{1\over \sigma^2}{\d\over \d\varrho} \left({\lambda(\varrho)^2\over 2}
\right)+\mu=0\eqno(11)$$
with $\lambda(\varrho)=1-{1 /\varrho}$. Note that  the solution of
this equation, $\lambda_s$, is such that $D\lambda_s$ is the function of
the combination $\mu\sigma^2$ only.
 In 1D case the solution of
 equation (11) can actually be obtained straightforwardly.
We will treat, however, this equation in a different manner, which will
be much more appropriate in 3D case.
We introduce the function $\mG(\tau)$,
$$\mG(\tau)={1\over 1+\tau}-1,\ \ \ \hbox{ so that }\
\varrho= \mG(-\lambda)+1.\eqno(12)$$

  After substitution of formula (12) into equation (11),
 the solution of equation (11), $\lambda_s$, is then given
by the implicit equation
$$D\lambda_s=\left.
-\mu\sigma^2{\d \mG\over \d \tau} \right\vert_{\tau=-D\lambda_s}.\eqno(13)$$
The generating function $C(\mu)$ of the cumulants is obtained by taking
 the logarithm
of the integral (10).
 In the low $\sigma$ limit it leads to retain only
the term that is under the exponential at the saddle point position.
We then obtain from (10),
$$\mC(\mu)=\mu\mG(-\tau_s)-{\tau_s^2\over 2\sigma^2},\ \ \
\tau_s=-D\lambda_s\eqno(14)$$
where the saddle point $\lambda_s$ (or equivalently $\tau_s$)
is given by the  equation (13).

The reason we  introduced  $\mG(\tau)$ is the following.
Equations  (13) and (14) have a structure of the Legendre transformation
from  $\mG(\tau)$ to $C(\mu)$
with the variable of the transformation equal to unity afterwards.
It  turns out that the function  $\mG(\tau)$ is the central object derived from
the basic dynamical equations in
the perturbation theory calculations  (Bernardeau 1992).
In the perturbation theory  $\mG(\tau)$ is defined as the generating
 function of the other averages -- vertices (see Sec. 4).
It is remarkable  that it corresponds
to the dynamics of the
``spherical''  collapse -- one dimensional in this case: $\mG(-\lambda)$ gives
the density contrast of a collapsing object of linear overdensity
$\lambda_0$, $\lambda =D(t)\,\lambda_0$, as function of time.
We will  discuss this derivation in general case in Sec. 4.
The reason   the Legendre transformation emerges is that two generating
 functions  $\mG(\tau)$ and $C(\mu)$ considered as the sum over the
 corresponding
 averages are connected through that transformation.
Hence, note that  in the 1D case
 expression (14) for $C(\mu)$ can be obtained  from the direct
 perturbation
series expansion of the basic equations as well.

The  crucial observation from
 the equation (14) is that the combination
$\mC(\mu)/\mu$ is a function of the combination  $\mu\sigma^2$ only. What
does it imply? The term in $\mu^p$, the coefficient of which is the
$p^{th}$ cumulant we are looking for, is then proportional to
 $\sigma^{2(p-1)}$.
It rigorously demonstrates a scaling relation between the cumulants
in the small $\sigma$ limit (i.e. at large scales).
The expression of these coefficients can be obtained with an expansion of the
function $\mC(\mu)$ with $\mu$:
$\mC(\mu)=\mu^2\sigma^2/2 +\mu^3\sigma^4+3\mu^4\sigma^6+...\ \ .$
 Then, for the skewness and kurtosis, for instance,  we get
$$S_3(0)=6,\eqno(15a)$$
and
$$S_4(0)=72.\eqno(15b)$$

These results can be extended by the further perturbative calculation
that takes into account the next terms of the expansion
in the steepest descent
method. It is then possible to get the first $\sigma$ corrections
for the expression of the skewness and the kurtosis. We then get
$$S_3(\sigma)=6+24\sigma^2+O(\sigma^4),\eqno(16a)$$
and
$$S_4(\sigma)=72+810\sigma^2+O(\sigma^4).\eqno(16b)$$
These expansions are not affected by the shape and parameters of
the cut-offs.
Theoretical curves (16) derived for small $\sigma$ are plotted
on Fig. 2, and indeed osculate the universal numerical curves for
small $\sigma$.

One of the  challenges of the study of the large-scale structure
formation is to find the $\sigma$ dependence of the $S_p$ parameters. From
the theoretical point of view, it is not clear
whether the $\sigma$ dependence can be accurately determined
with perturbation theory in the single stream approximation.
Multistreaming may change the behavior in an unknown way.
But we present  here the results in 1D case
to stress that perturbation theory does not prove at all
that the parameters $S_p$ are constant when $\sigma<1$.
The important property is that, for Gaussian initial fluctuations, they have
finite limits at small $\sigma$.
Note that the expansion of $S_p$
as a function of $\sigma$ similar to formula (16)  has never been done for
 the 3D case.

\vskip .5 cm
\leftline{\sl 2.4. The reconstruction method from the generating function
of the cumulants}

In the previous section we solved the problem how to find the
moments from the known PDF.
When the whole series of the cumulants is known it is possible to
to construct the density PDF from the
generating function of the cumulants by inversion
of the relation (10) (the inverse Laplace transformation):
$$P(\varrho)\d\varrho= {\d\varrho\over 2\pi\i}\int_{-\i\infty}^{+\i\infty}
\d \mu
\exp\left[-{C(\mu)}-{(\varrho-1)\mu}\right]
,\eqno(17)$$
where the integration is made in the complex plane.

In practice we can find the moments and generating function in the limit
of small $\sigma$.
For the further analysis
 let us define the function $\varphi(y)$
by the relation
$$\varphi(y, \sigma)=\sum_{p=1}^{\infty}S_p(\sigma){(-1)^{p-1}\over p!}y^p=
y-\sigma^2\mC(-y/\sigma^2), \eqno(18)$$
where we define $S_1=1,\ S_2=1$.
The advantage of this new function is that it is finite for an arbitrary
$y$ in the limit  of small $\sigma$.
This function is the generating function of the parameters $S_p(\sigma)$.
Then from (17) and (18) the density PDF is  given by
$$P(\varrho)\d\varrho={\d\varrho \over 2\pi\i\sigma^2}
\int_{-\i\infty}^{+\i\infty} \d y
\exp\left[-{\varphi(y, \sigma)\over\sigma^2}+
{\varrho y\over \sigma^2}\right],\eqno(19)$$
 Balian \& Schaeffer
(1989) used this form for the hierarchical ansatz, when there is no
 dependence on $\sigma$ in $\varphi(y, \sigma)$.
In the small $\sigma$ limit the $\sigma$ dependence that may
be contained in  $\varphi(y, \sigma)$ vanishes,
  $\varphi(y, \sigma) \to  \varphi(y)$,  and the results of
Balian and Schaeffer (1989) can then be used here, too.

Note that
general  formula (19) does not assume, a priori, that $\sigma$ is small
(as for the Edgeworth expansion in the next subsection).
It supposes, however, that the function $\varphi(y, \sigma)$
that is used is valid in the domain of application.
The reconstruction formula is of obvious interest when the function
 $\varphi(y, \sigma)$
can be derived from the first principles as it has been shown when
$\sigma=0$ in 1D case. We will see that it is also the
case  for small $\sigma$ limit  in 3D case.
 The reconstruction formula (19) is thus of general interest.

In Fig. 3 the density PDF in 1D case is calculated numerically with
the  reconstruction formula (19), where $\varphi(y)$ is calculated
from (14) and (18) in the small $\sigma$ approximation, i.e.
 ignoring $\sigma$-dependence in $\varphi(y,\sigma)$.
 It is then compared to the original
shape of the density PDF (Eq. [4]).
We find that the reconstruction method based on the generating function
works very well for the density interval $0 < \varrho \lsim 2$,
slightly worsening as $\sigma $ increases.
The shape of the high-density tail at $\varrho \gsim 2$
of the PDF from the reconstruction method differs from  that of the  actual
PDF.
 For the form (19), the density PDF
has an exponential  cutoff as $\exp[-\varrho/(x_*  \sigma^2)]$
 with $x_*=27/4=6.75$ (see Bernardeau 1992), which is very
different from the $\varrho^{-3}$ cutoff of formula (4).
 This
discrepancy is just due to the fact that we ignore  the $\sigma$-dependence
in  the parameters $S_p$.
 Therefore the discrepancy is increasing with  $\sigma$.
It is quite interesting that for moderate $\sigma$,  the $\sigma$-dependence
 affects the high-density tail
of PDF, but does not affect its shape around the maximum, nor
the low-density tail (see the Table  in Sec. 6).

\vskip .5 cm
\leftline{\sl 2.5. Reconstruction of  PDF from a few
 cumulants: The Edgeworth expansion}

In the previous section it has been shown how it is possible to derive
the shape of the density PDF out of the
generating function.
In this section we  report a method that can be used to
recover the shape of the PDF when only
a limited number  of
cumulants is known. In practice we may have only a few lowest cumulants,
 such as the skewness and the kurtosis.
In the case of the weakly non-linear dynamics, when slight departure from the
initial
Gaussian distribution is expected, one can use the general decomposition
 series around the Gaussian PDF
induced by the first non-zero cumulants. This decomposition
is known as the Gram-Charlier series (Kendall \& Stuart, 1958).
 Longuet-Higgins (1963) applied the Edgeworth form of the  Gram-Charlier
 series to the statistics of the
  weakly non-linearities
 in the theory of 2D sea waves. Inspired by this paper, we suggest
 to use  the Edgeworth's decomposition for the 1D and 3D cosmological
 density PDF (as reported in  Kofman 1993). We understand that similar ideas
were independently suggested by Juszkiewicz et al. (1993).

 The Edgeworth expansion can be derived from
  the form (19) of the density PDF.
Assuming that the density contrast  $\dta$
is  of the order of $\sigma$ and small, the relevant values
of  $y$ are also of the order of $\sigma$ and are thus assumed to be
small. It is then legitimate
to expand the function $\varphi(y)$
$$\varphi(y, \sigma)\approx
y-{1\over 2} y^2+{S_3\over 3!}\ y^3-{S_4\over 4!}\ y^4+{S_5\over 5!}\ y^5 \pm
\dots,\eqno(20)$$
where $S_p=S_p(\sigma)$.
To calculate the density PDF, we substitute the expansion
(20) into the  the integral (19). Then we make a further
expansion  of  the {\sl non-Gaussian} part of the factor
$ \exp\left[-{\varphi(y)/\sigma^2}\right]$
 with respect to both $y$ and $\sigma$
assuming they are of the same order, see Appendix B for details.

Finally we  obtain the so-called Edgeworth form of the Gram-Charlier
series for density PDF
$$
\eqalign{
P(\delta)\d\dta=&{1 \over (2\pi\sigma^2)^{1/2}}
\exp
\left( -\nu^2/2\right)
\biggl[1 + \sigma {S_3 \over 6} H_3\left(\nu\right)
+\sigma^2\biggl( {S_4 \over 24} H_4\left(\nu\right)
+{S_3^2\over72}H_6\left(\nu\right)
 \biggr)\biggr.\cr
&\biggl.+\sigma^3\biggl({S_5\over 120} H_5(\nu)+
{S_4 S_3\over 144} H_7(\nu)+ {S_3^3\over 1296} H_9(\nu)\biggr)
+ ... \biggr] \d\dta,\cr}\eqno(21)$$
where $H_n(\nu)$ are the Hermite polynoms (see Appendix B), $\nu=\dta/\sigma$.
This is a universal form for any slightly non-Gaussian dynamical models,
i.e. when $\sigma$ is small and $S_p$ are finite.
The actual forms of the parameters $S_p=S_p(\sigma)$ which depend
 on particular dynamics,  affect the  expansion (21)
with respect to $\sigma$.
Longuet-Higgins (1963), Juszkiewicz et al. (1993) and Kofman (1993) gave it
up to  $\sigma^2$-correction.
We give the Edgeworth expansion up to $\sigma^3$-correction,
for which the $\sigma$-dependence of $S_3$ (see, for instance, eq.[16a])
has to be taken into account.
The resulting approximate form (21) is a Gaussian distribution
multiplied by a corrective function -- in the square brackets --
 expanded with respect to
$\sigma$. The zero order  of this expansion gives simply a Gaussian
distribution; the first order  corrects it by taking into account the skewness;
the second order  by taking into account both the skewness and the kurtosis,
 etc.

Thus, it is  possible to get an approximate form of the density PDF
from  a few lowest known  cumulants.
This method is irreplaceable when only a few cumulants have been derived
from the first principles.
However, it is important to note that
this expansion has been made possible only for {\sl slightly }
non-Gaussian regime.
The validity domain of the form (21) is  limited to finite values
of $\dta/\sigma$.

Let us now apply the Edgeworth expansion (21) to the
 1D gravitational dynamics. We have to take into account
the $\sigma$ dependence of $S_3$ given in equation (16a),
and then obtain,
$$\eqalign{
P(\delta)\d\dta=&{1 \over (2\pi\sigma^2)^{1/2}}
\exp\left( -\nu^2/2\right)\biggl[1 + \sigma H_3\left(\nu\right)
+\sigma^2\biggl( 3 H_4\left(\nu\right)
+{1\over2}H_6\left(\nu\right) \biggr)\biggr.\cr
&\biggl.+\sigma^3\biggl( 11 H_5(\nu)+3 H_7(\nu)+{1\over 6}
H_9(\nu)+4 H_3(\nu)\biggr)+ ... \biggr] \d\dta.\cr}\eqno(22)$$

We plot the density PDF in 1D case reconstructed from the
Edgeworth expansion (22) in Fig. 4, and compare it with the
actual PDF (4), for $\sigma=0.1,$ and $\sigma=0.3$.
We can see that a few iterations of the expansion (22) reproduce
  the peak
of $ P(\delta)$  in the interval $\vert \delta \vert \lsim 0.5$ around it
for   small $\sigma$ relatively well.
It reproduces well  the shift of the maximum towards
the low density.
 It completely fails to reproduce  $ P(\delta)$
outside of this interval where spurious wiggles appear.
For a given value of $\sigma$,  each next
$\sigma$ iteration quite slowly  improves the approximation.
Unfortunately, the method is rapidly worsening as  $\sigma$ increases,
and in practice is useless for  $\sigma \gsim 0.5$.

The usual  measurements of  the lowest cumulants are significantly
affected by  the high density tail of the PDF, i.e. the rare events.
It is interesting to note, that in the context of the reconstruction
methods, the lowest
cumulants alone are responsible for the shift of the peak of
$P(\delta)$. Therefore the measurement of the shape of the PDF maximum,
which statistically is more robust,
can provide an alternative method of evaluation of the lowest cumulants.

%\page
\bigskip\bigskip
\centerline{\bf 3. THE 3D DYNAMICS WITH THE ZEL'DOVICH APPROXIMATION}
\medskip

We consider here the  statistics of the cosmic density field in 3D Zel'dovich
 approximation
in a similar manner as it was done in 1D case.
 The most
important change is that the Zel'dovich solution no longer reproduces
the exact  dynamics but is an approximation.
It is, however, thought to be a good description, so that it is worth
investigating the statistical properties that can be obtained in
 this approximation. We report here a different derivation of
 what have been done by Kofman (1991) and Kofman \etal (1994) for
Gaussian initial fluctuations. The new method
(Kofman 1994) also allows
  to extend the results to non-Gaussian  adiabatic initial
 fluctuations.

\vskip .5 cm
\leftline{\sl 3.1. Zel'dovich approximation for filtered initial fluctuations}

For the sake of  simplicity we assume that the universe is filled by the
 pressureless matter.
The growing mode of adiabatic perturbations is  $D(t)$.
Let $\vx,  \vv =a\,{\d\vx / \d t},  \rho  (t, \vx)$ and
$\phi (t,\vx)$ be, respectively, the comoving coordinates, peculiar velocity,
 density of dark matter and peculiar Newtonian
gravitational potential.
It is convenient to introduce a new time variable, $D(t)$, then a comoving
velocity is $\vu ={\d\vx / \d D}$.
 The growing mode is
non-rotational, so that the velocity field is potential.
Let $\Phi$ be the velocity potential so that $\vu = \nabla_{\rm x}\Phi$.

The gravitational dynamics of the cosmological system
 is  complicated and requires the N-body
  simulations.
For interesting cosmological models such as the CDM scenario, the structure
formation looks like complicated hierarchical pancaking and clustering from
very small to large cosmological scales (e.g., see
  Shandarin \& Zel'dovich 1989, Kofman et al. 1992).
However the gravitational clustering at  sufficiently large  scales
  $R$  can be
considered  in the quasilinear theory  in a single stream regime
ignoring small scale details.
   For this goal we  use the Zel'dovich approximation  but apply it for the
filtered  initial gravitational potential.
This approach, sometimes called the truncated Zel'dovich approximation,
was used for different purposes  in papers
Bond and Couchman (1986), Kofman (1991),
Shandarin (1992), Kofman et al. (1992), Coles et al. (1993),
Kofman et al. (1994). From a mathematical point of view the
truncated Zel'dovich approximation
means just that the Zel'dovich approximation is applied
to the truncated initial potential  to ensure  being
 in the single stream quasilinear regime.

To describe the motion of particles we can introduce the
 tensor of the velocity derivatives $S_{ij}(\vx, t)=-\nabla_{\rm x_i} \vv_j$
 in the Eulerian space;
 for the potential motion it is reduced to
$S_{ij}=-\nabla_{\rm x_i}\nabla_{\rm x_j}\Phi$.
Let $\lambda_i$ be its  eigenvalues.
The field of the  $S_{ij}(t)$-tensor evolves in time, its initial
value (in the Lagrangian space)
 coincides with the Lagrangian  deformation tensor
$D_{ij}=-\nabla_{\rm q_i}\nabla_{\rm q_j}\Phi_0$. From dynamical equations
one can easily show that the eigenvalues
  $\lambda_i(t)$ are related to their initial values $\lambda_{0i}$, by
$\lambda_i(t)={\lambda_{0i}/({1+D(t)\lambda_{0i}})}.$
The local density can  be obtained  by the inverse
of the Jacobian of the transformation between $\vq$
and $\vx$, so that
$$\rhos   =  {{\varrho_0}
 \over {\vert(1-D\lambda_{01})(1-D\lambda_{02})(1-D\lambda_{03})\vert}}.
\eqno(23)$$
In the ZA the statistics of the evolved field can then entirely be
obtained by the statistics of the initial local density $\varrho_0$
and the initial eigenvalues $\lambda_0$-s. For adiabatical perturbations
 $\varrho_0=1$.

\vskip .5 cm
\leftline{\sl 3.2. Joint PDF in the ZA for arbitrary initial statistics}

In the last section 3.1 all  relationships were obtained without making
 any assumption
on the initial statistics. The cosmic density PDF
 can then be obtained
from the initial joint PDF of all involved cosmic fields:
$W_0(\varrho_0,\lambda_{01},\lambda_{02},\lambda_{03},
 \vec u_0, \Phi_0)\
\d\varrho_0\ \d\lambda_{01}\ \d\lambda_{02}\ \d\lambda_{03}\
\d^3u_0\ \d\Phi_0$.
That the statistics can be completely made from the statistical
behavior of all variables is entirely
due to the use of the Zel'dovich approximation. If this
approximation were released it would no longer be true.

The density PDF can be obtained in general case of arbitrary initial
condition by integrating the combination
$ \delta \bigl[ \rhos
\vert (1-D\lambda_{01})(1-D\lambda_{02})(1-D\lambda_{03})\vert-  1\bigr] \times
 W_0(\varrho_0,\lambda_{01},\lambda_{02},\lambda_{03},
 \vec u_0, \Phi_0)$ over all involved variables except density.
This is one of the  new results of our paper. Density PDF for non-Gaussian
initial fluctuations will be considered in a separate paper (Kofman 1994).

\vskip .5 cm
\leftline{\sl 3.3. Density PDF for Gaussian initial fluctuations}

In this Section we study  the statistics of the continuous cosmological
fields evolving from the initial Gaussian fluctuations, which
is the general frame for most cosmological models.
For the  Gaussian initial conditions we can omit
$\vec u_0$ and $\Phi_0$ in the initial joint PDF and write it  as
$$
 W_0( \rhos_0, \lambda_{01}, \lambda_{02}, \lambda_{03} )
\d \varrho_0\ \d\lambda_{01}\ \d\lambda_{02}\ \d\lambda_{03}=
 \delta (\rhos_0-1) \ \d\varrho_0\
 M_0( \lambda_{01}, \lambda_{02}, \lambda_{03})
\ \d\lambda_{01}\ \d\lambda_{02}\ \d\lambda_{03}.\eqno(24)
$$
 The first factor -- the Dirac $\delta$-function -- is the initial
density distribution function, which corresponds to the perfectly
homogeneous density distribution $\rhos_0=1$.  This is just
the formal limit of the Gaussian density distribution with
$\sigma \to 0$. This factorization is expected to take
place for cosmological models with small adiabatic initial fluctuations.
The second factor is the joint distribution function
of the eigenvalues of the initial deformation tensor for
an initial Gaussian displacement field (Doroshkevich 1970)
$$\eqalign{
\d\lambda_{01}\ \d\lambda_{02}\ \d\lambda_{03}&
\ M_0(\lam_{01},\lam_{02},\lam_{03})=
\ \d\lambda_{01}\ \d\lambda_{02}\ \d\lambda_{03}\cr
&\times {5^{5/2}\, 27 \over 8 \pi \sigma_0^6}\
(\lam_{01}-\lam_{02})(\lam_{01}-\lam_{03})(\lam_{02}-\lam_{03})\
{\rm exp}\left[-{1\over\sigma_0^2}
\biggl(3J_{01}^2-{15 \over 2} J_{02} \biggr)\right] \
,\cr} \eqno(25)
$$
where $J_{01}=\lam_{01} +\lam_{02}+ \lam_{03}$ and
 $ J_{02}=\lam_{01}\lam_{02} +\lam_{01}\lam_{03} + \lam_{02}\lam_{03}$.

The shape of the density PDF can then be obtained by the
change of variable $\varrho_0$ to $\varrho$
and by integrating over $\lambda_0$-s in (24).
We then have
$$\eqalign{
P(\rhos, D)\d\rhos
 = \d\rhos&
\int  \d \lambda_{01}\ \d \lambda_{02}\ \d \lambda_{03}\
\delta \bigl[ \rhos
\vert (1-D\lambda_{01})(1-D\lambda_{02})(1-D\lambda_{03})\vert-  1\bigr]\cr
&\times M_0 (\lambda_{01},\lambda_{02},\lambda_{03})
.\cr}
\eqno(26)
$$
We get the formal expression for the Eulerian  density PDF in the ZA.

Substituting the expression (25) into the integral (26),
after some tedious algebra, we can reduce the integral (26) to a
simpler one-dimensional  integral which has to be evaluated
numerically,
$$\eqalign{
P(\rhos, D) \d \rhos=&
 {{9 \cdot 5^{3/2} \ \d\rhos} \over {4\pi N_s \rhos^3 \sigma^4}}
\int_{3 ({\bar \rhos \over \rhos })^{1/3}} ^{\infty}  d s\
e^{-{(s-3)^2 / 2 \sigma^2}}\cr
&\times
\left( 1+ e^{-{6s/ \sigma^2}} \right)
\ \left( e^{-{\beta_1^2 / 2\sigma^2}}
   +e^{-{\beta_2^2 / 2\sigma^2}}
   -e^{-{\beta_3^2 / 2\sigma^2}}  \right)\ ,\cr}\eqno(27)
$$
$$
\beta_n (s) \equiv s \cdot \,5^{1/2} \left( {1\over2}
+\cos\left[{2\over3}(n-1)\pi
+{1\over3} \arccos \left({{54{ \bar \rhos}^3} \over \rhos s^3}
-1 \right)\right]\right) ,
$$
where the parameter $\sigma(t)=D(t)\sigma_{0}$ is
 the standard deviation of the density
fluctuations
$\rhos/\bar\rhos$ in the  linear theory, and $N_s$ is the mean number
 of streams, $N_s=1$ in the single stream regime.
The expression (27) was derived by a different
method  earlier on
by Kofman (1991), Kofman et al. (1994).

This 3D formula corresponds  to the formula (4) derived in 1D.
The analytical expression is obviously different, but the method to
build it follows the same scheme: the density PDF is obtained
from the joint PDF of the eigenvalues $\lambda_0$ of the initial
local deformation  tensor.
In the limit of very small $\sigma$ the formula (27) is reduced to
the Gaussian distribution.
 We plot the PDF calculated numerically from formula (27) in Fig. 5.
For moderate values of $\sigma$ density PDF calculated in the ZA
is in good agreement with the PDF from N-body CDM simulations,
see Kofman et al. (1994), and also Fig. 7. One can expect formula (27)
works even better for those models, for which the pancaking is more
pronounced.

%\page
\vskip .5 cm
\leftline{\sl 3.4. Calculation of $S_p$ for small $\sigma$ in the ZA}

The 3D density PDF in the ZA also has the caustic-induced
 $\varrho^{-3}$-asymptota. As a result, the moments cannot be formally
defined in the ZA for any given value of  $\sigma$.
However, we can
regularize the PDF in the Zel'dovich approximation
 by  cutting off the  high density tail as we did
in Sec. 2.2. for 1D case. Then we are
 able to define the cumulants
and derive their  large scale properties.
We use the same two regularizations, the sharp cutoff and the exponential
 cutoff
 as for the 1D case, see equation (4).  The parameters
 $S_3$ and $S_4$, defined as in  relation (7)
and calculated with numerical integrations of the moments, are plotted
in Fig. 6  as functions of $\sigma$.
They exhibit a very similar qualitative behavior than for the 1D case.
Numerical shapes of $S_p(\sigma)$ are universal for small $\sigma \lsim 0.3$,
i.e. independent of the form of the cutoff and on the parameter of truncation.
However, their  quantities in the small $\sigma$ limit
 are changed compared with 1D case. In 3D case
we found  $S_3 \approx 4$ and $S_4\approx  30$.
For $\sigma \gsim 0.3$, $S_3$ and $S_4$ depend on the shape of the
regularization functions,
therefore the moments in the ZA for moderate and large $\sigma$
are poor-defined.

In the small $\sigma$ limit more advanced analytical progress can be done
for the Gaussian initial fluctuation.
Grinstein \& Wise (1987), Munshi \& Starobinsky (1993) calculated $S_3(0)=4$;
Bernardeau \etal (1994), Catelan \& Moscardini (1993) calculated
$S_3(0)$ and $S_4(0)=272/9$, all  from the perturbation theory around the ZA.
The new result we present in the rest of this section
is the analytical  derivation of the generating function of the cumulants,
and consequently, in principle,  {\sl all} the cumulants themselves, in the ZA.

The most straightforward method of the calculation of the generating
function $C(\mu)$ and   cumulants is
 the same as for the 1D case.
Let us use the
 integral (10) for $C(\mu)$,
where for the density PDF in the ZA we use the integral (26), which
includes the joint PDF
  of the eigenvalues
$\lambda_{0i}$ and the $\delta$-function of the density.
Integrating   the $\delta$-function over the density,
 we get the integral over $\lambda_{0i}$-s
$$\eqalign{
\exp \mC(\mu)=&{5^{5/2}\, 27 \over 8 \pi \sigma_0^6}
\int{\d \lambda_{01}\ \d\lambda_{02}\ \d\lambda_{03}\over
 \vert(1-D\lambda_{01})(1-D\lambda_{02})(1-D\lambda_{03})\vert\ }
(\lam_{01}-\lam_{02})(\lam_{01}-\lam_{03})(\lam_{02}-\lam_{03})\cr
&\times{\rm exp}\left[-{1\over\sigma_0^2}
\biggl( 3J_{01}^2-{15 \over 2} J_{02} \biggr)
+\mu
\left({1\over \vert(1-D\lambda_{01})(1-D\lambda_{02})(1-D\lambda_{03})\vert }
-1\right) \right].\cr}
  \eqno(28)$$

In the limit of small $\sigma$ we can apply the steepest descent method to
evaluate this integral. This is a triple integral that, however, can be
calculated since
$\lam_{0i}$-s are involved in the integrand in symmetric  combinations only.
There are three equations for the saddle point
$(\lam_{s1}, \lam_{s2}, \lam_{s3})$ in the 3D $\lam_{i}$-space, which admit
a simple symmetric solution
$$\eqalign{
\lambda_{01}&=\lambda_{02}=\lambda_{03}=\lambda_{s}\,, \cr
3D\lambda_{s}&={\sigma^2\mu\over (1-D\lambda_{s})^4}. \cr}\eqno(29)$$
Note that $D\lambda_s$ is a function of the combination $\sigma^2\mu$
only, as it was in 1D case.
This fifth-order equation cannot be solved analytically.
Despite  that, the introduction of the machinery of $\mG(\tau)$-function
described in Sec. 2.3 helps to overcome the problem.

 The saddle point equation (29) can be rewritten in terms of
$\mG^Z(\tau)$-function. Indeed, the equation
$$\tau_s=\mu\sigma^2 {\d \mG^Z \over
\d \tau}(\tau_s),\ \ \ \ \tau_s\equiv -3D\lambda_s,\eqno(30)$$
is reduced to the algebraic equation (29), if
we choose
$$\mG^Z(\tau)={1\over (1+\tau/3)^3}-1 .\eqno(31)$$
Then the leading factor of the integral (28)
for small $\sigma$, emerging from the steepest descent method,  gives us
the expression for  $\mC(\mu)$:
$$\mC^Z(\mu)=\mu\mG^Z(\tau_s)-{\tau_s^2\over 2\sigma^2},\eqno(32a)$$
or equivalently
$$\varphi^Z(y)=y+y\mG^Z(\tau_s)+{\tau_s^2\over 2},\ \
\tau_s=-y{\d\over\d\tau}\mG^Z(\tau).\eqno(32b)$$

The formulae (31), (32) are  the 3D counterparts  of (12) and (14)
 for the 1D case.
Again, equations  (32) have the structure of the Legendre transformation
from  $\mG^Z(\tau)$ to $C^Z(\mu)$
with the variable of the transformation equal to unity afterwards.
It is remarkable, that the formulae (31) and (32) can  be  derived
 independently
from the perturbation series based on the dynamical equations
of  ZA, similar to the method of  Bernardeau (1992)
based on the perturbation series of cosmological equations (see also Sec. 4),
and what we discussed in Sec. 2.
With this approach we obtain
form (32) for $\mC^Z(\mu)$ where the function
$\mG^Z(\tau)$ describes the collapse of the density contrast  of a symmetric
perturbation of linear overdensity $-\tau$ in the equations corresponding to
the three dimensional
 ZA. The solution of this problem is given by formula (31).
Note, that the form (31) has been derived here by a totally different method!

Using the expression (32), we can derive analytically the
full series of the $S_p(0)$ parameters in the ZA.
For instance we have
$$\eqalignno{
S_3^{Z}(0)&=4,&(33a)\cr
S_4^{Z}(0)&={272\over9} \approx 30,&(33b)\cr
S_5^{Z}(0)&={3080 \over 9 } \approx 342.&(33c)\cr}
$$
The first two figures  are represented by circles in Fig. 6,
and give the small $\sigma$ limit of the numerical curves $S_3(\sigma)$ and
 $S_4(\sigma)$.
The next terms of the asymptotic series of the integral (28)
 would allow, in principle, to derive analytically the $\sigma$
expansion of the  parameters $S_p(\sigma)$, similar to
1D decompositions (16). We leave this exercise out of the scope of the paper.
It is interesting to note that the skewness, kurtosis, etc,
in the 3D case are systematically smaller then corresponding numbers of the
 1D case, c.f. (33) and (15). As numerical curves of Figs. 2 and 6
show, the $\sigma$ dependence (where it is well-defined) is also
weaker in 3D case than in 1D case. In the ZA the
departure from the Gaussian distribution is than
faster in 1D case then in 3D case.

\vskip .5 cm
\leftline{\sl 3.5  Effects of the  final smoothing in the ZA for
 small $\sigma$}

In the previous section we found the parameters
 $S_p$ and
 the generating function in the limit of small $\sigma$
 in the framework of the ZA. Using the machinery
of the reconstruction, described in Sec. 2.,  we can
obtain the density PDF in this limit from the generating function (19), or
 from the
Edgeworth asymptotic expansion (21). However, it has  little sense since
we know the PDF (27) for an arbitrary $\sigma$ in the ZA. What is of greater
interest is
to use the reconstruction technics when the
final smoothing effects in the generating function and $S_p$ are
taken into account.
It can be done for small $\sigma$ only, and allows, in principle,
to reconstruct
the density PDF in the ZA with final smoothing.

To get the  PDF for the continuous field for
 different $\sigma$, we need to filter the Eulerian density field,
either in observational surveys,  N-body simulations, or analytical
approximations.
 The truncated Zel'dovich approximation,
 introduced in Sec. 3.1, is based on the smoothing of the initial fluctuations.
We
recall, that the initial smoothing is  inevitable anyway, otherwise
we get the multiple streaming regime, for which the ZA is not applicable at
 all.  The fact that the order of the smoothing and the
dynamical evolution  not commute was noticed by Kofman et al. (1994).
It is quite complicated to incorporate analytical approaches given
in that paper or in Sec. 3,
with the final smoothing. The reason is that the smoothed density does not only
depends on the local eigenvalues of the deformation tensor at one
point, but on the  behavior of the deformation tensor
in  a whole smoothing  large area.

When the filtering is taken into account, the PDF is expected
to slightly depend on the shape of the power spectrum,
since it is known after Goroff \etal (1986)
that the parameters $S_p$ depend on it.
On the other hand, the PDF (27) does not have any dependence
with the power spectrum and is only defined by the value of the
rms density fluctuation $\sigma$, showing some limitations for the
practical use of this result.
For the sake of simplicity, we will assume
in the following, that the dependence
on the power spectrum  can be reduced to the dependence
on the effective index $n$ at the scale at which the field is filtered.
The index $n$ is the logarithmic derivative of $\sigma$ with the scale $R$.

There was an attempt by Padmanabhan and Subramanian (1993)
to take into account the additional, final smoothing. They made an additional
approximation within the ZA, assuming that the typical collapse is spherical.
Unfortunately this assumption is valid for a small fraction of the Lagrangian
volume, and the resulting distribution poorly fit the numerical PDF.
Juszkiewicz et al. (1993) suggested an interesting phenomenological conjecture
in order to take into account the final smoothing effects. They
 pointed out
that the $\sigma$ and $n$ dependence of the density
PDF (for small $\sigma$) is mainly contained in the $S_3\,\sigma$ combination.
In the context of the ZA  this property leads to the substitution
$\sigma \to \sigma\,S_3(n)/4 $ in the density PDF (27).

In this Section we present the generalization (in the limit
of small $\sigma$) of the generating function (32) in
the ZA that takes the final smoothing into account.
The top-hat filtering is the simplest model of smoothing for which
a complete analytical study can be done.
 We report the results
of a general method developed by Bernardeau (1994)
and applied here for the ZA.
The basic idea is that the non-linear contributions
to the $S_p$ parameters are not affected by the filtering at a given mass $M_0$
scale (top-hat filtering in the Lagrangian space). This property is valid
for the ZA as well as for  the exact single stream cosmological dynamics,
 but may not be
true for other dynamical approximations.
The filtering effects at a given radius $R_0$
can then be calculated from a transformation of the density PDF
from Lagrangian space to Eulerian space:
$\int^{\infty}_{\rhos_0} \rhos P_E(\rhos, R_0)\d \rhos=
\int^{\infty}_{\rhos_0} P_L(\rhos, M_0)\d \rhos$, where $M_0=4\pi R_0^3/3$
(see  Bernardeau 1994 for details).
 It is striking to
see that the result  in terms of a $\mG$-function can be derived from that
through (32) and the reconstruction formula (19).
 For the ZA with final smoothing
this function
 will be denoted as $\mG^{ZS}$.
The mathematical transformation
to obtain $\mG^{ZS}$ from $\mG^{Z}$ (corresponding to the case without
smoothing) is then given by the formula,
$$
\eqalign{
\mG^{ZS}(\tau)=\mG^Z\left\{\tau\ {\sigma\left([1+\mG^{ZS}
(\tau_s)]^{1/3}R_0\right)\over\sigma(R_0)}\right\}
,\cr}\eqno(34)$$
where $R_0$ is the filtering scale and
$\sigma(R_0)$ is the rms density fluctuation at this scale.
The generating function of the cumulants is then given by
$$\eqalign{
\varphi^{ZS}(y)&=y+y\mG^{ZS}(\tau)+{1\over 2}\tau^2,\ \
\tau=-y{\d \over \d \tau}\mG^{ZS}(\tau).\cr}\eqno(35)
$$

For instance the coefficients $S^{ZS}_3(0)$ and $S^{ZS}_4(0)$,
 for a power law spectrum of index $n$, are
$$\eqalign{
S^{ZS}_3(0)&={4}-(n+3),\cr
S^{ZS}_4(0)&={272 \over 9}-{50\over3}(n+3)+{7\over3}(n+3)^2.\cr}\eqno(36)$$
The smoothing correction given by formulae (34), (35) are quite complicated,
but final formulae (36) are rather readable.

Fortunately, for one of the most intersting cases $n=-1$,
corresponding to
the CDM power spectrum on the galaxy clustering scale, there is a simple
solution of equation (34)
$$\mG^{ZS}(\tau)=(1-\tau/3)^{3}-1,\eqno(37)$$
note difference with eq. (31) for  $\mG^{Z}(\tau)$.
Using this simple form, it is  possible then   in case $n=-1$ to build the PDF
 from the reconstruction
formula (19).
 We present the result for PDF for this interesting
 particular case in Fig. 5 and compare it to the
shape of the density PDF in the absence of smoothing obtained both
with the formula (27) and  with the reconstruction formula (19).
As in 1D, the reconstruction method (that neglects the $\sigma$-dependence
of the $S_p$ parameters) leads to a less extended
 high density (exponential) tail,
but does not change very much the low density behavior.
When the filtering effects are taken into account the density PDF, however,
tends to have weaker non-Gaussian features both at the
high- and  low-density tail.
The recipe to build density PDF in ZA with smoothing for arbitrary
 $n$ is consistent in substituting eqs.(34), (31) into the general
reconstruction formula (19), but valid for small $\sigma$ only.

\bigskip
\leftline{\bf 4. SUMMARIZING PERTURBATION SERIES FROM EXACT DYNAMICS}

In the previous section we calculated the statistics using
the Zel'dovich approximation. Beyond the Zel'dovich approximation,
 that is when one assumes only  the single
stream regime, the general form of the density
PDF, that would be the counterpart of (27) in ZA, has never been obtained.
However, using perturbation theory  for Gaussian initial conditions,
it is  possible to derive
cumulants of the density PDF. The approximation which is admitted
to apply the perturbation theory, is that the gravitational clustering
at sufficiently large scales can be considered in the single stream regime
ignoring small scale details.
This approach to derive cumulants in quasi-linear dynamics has been
intensively used in the literature (Peebles 1980, Fry 1984, Goroff \etal 1986,
Bouchet \etal 1992, Juszkiewicz \etal 1993, etc.).
 Within this approximation to the complicated
actual dynamics, the closed form for the
generating function of the cumulants is rigorously  derived in the
limit of small $\sigma$ (Bernardeau 1992, 1994). Since the cumulants
very slowly increase with $\sigma$ while $\sigma <1$, at least for
$n\approx-1$,
 these results can be extrapolated
across the whole quasilinear regime, which makes them very useful.
In this section we first recall the results of calculation of $S_p(0)$ and the
 PDF
based on the perturbation theory without final smoothing, as it was
derived in Bernardeau (1992). Then we report the corresponding results when
final smoothing is included (Bernardeau 1994),  which are the
 most important for practical applications. Then we show how the general
method of the Edgeworth expansion can be used with the cosmological
dynamical perturbation theory.

\vskip .5 cm
\leftline{\sl 4.1. Reconstruction of PDF through  $S_p$ in the
  small $\sigma$ limit from the exact dynamics   }

In 1D,
the large--scale limit (e.g. [15]) of the parameters $S_p$ is exact.
 In 3D  the Zel'dovich solution
 is really an approximation, at any stage of the dynamics, so that
the large--scale limit of the parameters $S^{Z}_p(0)$ from the ZA
is also an approximation to the exact $S_p(0)$.
However, it is possible to calculate
the  values of $S_p(0)$ from the exact single stream cosmological dynamics,
without having to know  the shape of the density PDF a priori.

The principle of the calculation has been given by
Bernardeau (1992) and is summarized in Appendix C.
In this section we recall the results that have been obtained
to compare them with those obtained from other methods.
One can use the basic single stream cosmological equation of gravitational
instability, and seek the solution in the form of perturbation series,
e.g. $\delta =\sum_{p=1}^{\infty}\delta^{(p)}$, see Appendix C.
Then there is a set of basic equations in each order $p$ for $\delta^{(p)}$.
Let us define the   following connected (normalized) averages:
 $\nu_p \equiv \mg\delta^{(p)} (\delta^{(1)})^p\md_c/\sigma^{2p}$. Next, we
   construct
the generating function of these averages
$$
\mG(\tau)=\sum_{p=1}^{\infty} {\nu_p \over p!}(-\tau)^p.
$$
The choice of that particular combination is based on the observation that
there is a closed equation for $\mG(\tau)$. Indeed,
multiplying equations  for  $\delta^{(p)}$ by $(\delta^{(1)})^p$ and averaging
the result, and then summarizing the hierarchy of equations, one can
derive the single equation for $\mG(\tau)$ (eq. [C7]).
It is remarkable that this equation does not contain space derivatives,
and has {\it exactly} the same form as the equation of the spherically
symmetric collapse of the ``overdense'' $\mG(\tau)$ with a ``scalar factor''
$\tau$.
A good approximate analytical solution of the ``spherical collapse'' is
 given by
$$\mG(\tau) \approx {1\over ( 1+\tau/1.5)^{1.5}}-1.\eqno(38)$$
The function $\mG(\tau)$, in principle, depends on the
cosmological parameters, but  weakly.
The closed form (38) is actually a fitting formula
that turns out to be extremely accurate for any cosmological
parameters.

The generating function $\varphi(y)$ of the $S_p(0)$ parameters (18)
 can then be built
with the generating function (38) of $\nu_p$ parameters.
 In Field Theory
the averages like  $ \mg\delta^{(p)} (\delta^{(1)})^p\md_c$  are known as
 {\it vertices},  the averages like the
  cumulants  $\mg\delta^{(p)}\md_c$ then are {\it trees}.
They can be represented by diagrams where factor  $\sigma$ corresponds to
the lines, and factor  $\nu_p$ to
 the vertices (for a vertex connecting $p$ lines, see
Appendix C for details).
There is a very useful and deep general result which links the generating
functions
of   vertices  and cumulants:
 the generating function of cumulants  is connected to the generating function
of vertices
 through the Legendre transformation (see Bernardeau \& Schaeffer 1992
for references). In the cosmological context the Legendre
 transformation reads as
$$\eqalign{
\varphi(y)&=y+y\mG(\tau)+{1\over 2}\tau^2,\ \ \
\tau=-y\mG'(\tau).\cr}\eqno(39)
$$

Now using (39) and (20), one can derive the whole series of the $S_p(0)$
parameters.
For instance, we have
$$\eqalign{
S_3(0)&={34\over7} \approx 4.9,\cr
S_4(0)&={60712\over 1323} \approx 45.9.\cr}\eqno(40)
$$

This approach allows to derive the parameters $S_p$ for $\sigma=0$.
Direct perturbative calculations also allow to obtain the parameters
$S_3(0)$, $S_4(0)$ (Peebles 1980, Fry 1984)
and, additionally,  should  admit  derivation of
 their further $\sigma$ dependence.
Unfortunately, no $\sigma$ corrections of $S_p(\sigma)$ were
obtained from the perturbation series so far, but
analysis of the numerical simulations indicates that $\sigma$ dependence
of $S_p(\sigma)$ is rather weak in the quasilinear regime
(Juszkiewicz et al 1993, and hereafter \S 5), at least for the
interesting cases $n=0, -1$.
Outside of the quasilinear regime,
   $S_p(\sigma)$ dependence might
be noticeable, as some numerical simulations indicate (Lucchin \etal 1994).
Note, however, possible impacts of the numerical effects on these calculations
(Colombi \etal 1993).

We can reconstruct the density PDF by substituting (39) in
the general formula (19).
 The resulting shape of the density
PDF is presented in Fig. 7. Note that
this PDF depends on $\sigma$ only and not on the power spectrum,
as it was  in the ZA without smoothing.

\vskip .5 cm
\leftline{\sl 4.2. The final  smoothing in $S_p$ and PDF for small $\sigma$}

It is also possible
to get the full series of the $S^S_p(0)$
parameters when the final smoothing, with a top-hat window function,
is taken into account.
We can do it via  the same method  we used earlier  for the
ZA in Sec. 3.6.
 The generating function of the cumulants is
defined via a $\mG^S$--function given by the relationship,
$$
\eqalign{
\mG^{S}(\tau)=\mG\left\{\tau{\sigma\left([1+\mG^S
(\tau)]^{1/3}R_0\right)\over\sigma(R_0)}\right\}
.\cr}\eqno(41)$$
where $R_0$ is the filtering scale and
$\sigma(R_0)$ is the rms density fluctuation at this scale.
The generating function of the cumulants is then given by
$$\eqalign{
\varphi^S(y)&=y+y\mG^S(\tau)+{1\over 2}\tau^2,\ \ \
\tau=-y{\d \over \d \tau}\mG^S(\tau).\cr}\eqno(42)
$$
For instance the coefficients $S^S_3(0)$ and $S^S_4(0)$,
 for a power law spectrum of index $n$, are
$$\eqalign{
S^S_3(0)&={34\over 7}-(n+3),\cr
S^S_4(0)&={60712\over1323}-{62\over3}(n+3)+{7\over3}(n+3)^2.\cr}\eqno(43)$$
We derive (43) from the generating function (41). The same coefficients
$S_3^S(0)$ and $S_4^S(0)$ as function of $n$ were derived independently
directly  from the
perturbation theory (Juszkiewicz \etal 1993, Bernardeau 1993), which confirms
the validity of the general formula (41).

Substituting
(42) into (19), we  can reconstruct the density PDF under all three
assumptions. Now it also depends on the power spectrum.
The properties of the resulting density PDF are given in details
by Bernardeau (1994), and are recalled in the Table  of Sec. 6 below.

Despite the fact that the $\sigma$ dependence on the $S_p$
parameters has been neglected, it
seems that the formula (19), with the generating function from (42), is
the most reliable analytical expression for the density PDF in the mildly
non-linear regime (see Sec. 5) for $n\gsim -2$.

\vskip .5 cm
\leftline{\sl 4.3. The Edgeworth expansion in 3D case}

The accuracy of the Edgeworth
decomposition in a realistic case
of 3D exact single stream dynamics with final smoothing is worth examining.
 Using the expressions of $S_3$
and $S_4$ (eq. [43]) we can use the decomposition (22) up to the
second order in $\sigma$. Note that the use of this decomposition
up to the third order would require the determination of the
first $\sigma$-correction  $\sim \sigma^2$ in $S_3$ which we do not know yet.

In Fig. 8 we plot the density PDF in 3D case reconstructed from the
Edgeworth expansion (22) with $S_p(0)$ from (43) for $n=-1$,
 and compare it with the
 PDF from (42)  and (19),  for $\sigma=0.3$ and $\sigma=0.5$.
We can see that a couple of iterations of the expansion (22) reproduce
 the peak
of $ P(\delta)$  in the interval $\vert \delta \vert \lsim 0.5$ around it
for small $\sigma$ relatively well.
It reproduces well  the shift of the maximum towards
the low density.
 It  fails to reproduce  $ P(\delta)$
outside of this interval.
For a given value of $\sigma$,  each next
$\sigma$ iteration improves the approximation  quite slowly.
The reconstruction is rapidly worsening as  $\sigma$ increases,
and in practice is useless for  $\sigma \gsim 0.5$.

As the $S_3$ and $S_4$ parameters are lower than in 1D,
the accuracy of the decomposition in 3D case is better.
Correspondingly, in 3D case with the final smoothing, the accuracy
of the Edgeworth decomposition is also better for smaller $S_3$ and $S_4$,
i.e. for larger index $n$.
Additionally,  from the comparison of 1D and  3D cases, one can infer that
the the Edgeworth decomposition is of interest when the actual skewness
of the distribution function is small, that is when $S_3 \, \sigma$
is small.

%\page
\vskip 1 cm
\leftline{\bf 5. COMPARISON WITH PDF AND $S_p$ FROM N-BODY SIMULATIONS}

We have derived analytically two forms (27) and (19) of density PDF in two
reasonable dynamical approximations, the ZA and the perturbation theory
 in a single
stream regime.
Apparently, there is no universal formula for the PDF in general.
Now we compare our analytical results with those from cosmological N-body
simulations.
The distribution function (27) has already been
tested in a previous work, Kofman \etal (1994) with
a density field that had been filtered with a Gaussian window function.
 From the theoretical point of view, the effect of filtering is
 better known for a top--hat
window function. We therefore run  a new series of tests with the top-hat
smoothing.

We used a large numerical simulation kindly
provided by Couchman (Couchman 1991). The simulation has been
made in a box of 200 $h^{-1}$Mpc size with periodic boundary
conditions, and contains $2.1\times 10^{6}$ particles. It used
an adaptive P$^3$M code and the initial conditions correspond to a CDM
power spectrum with $\Omega=1$, $H_0=50$km s$^{-1}$ Mpc$^{-1}$ and the
 bias parameter $b \approx 1.0$. We made three filterings with a top--hat
window function at two different time steps, at redshifts $z=0.6$ and $z=0$.
 The three different filtering radii we choose were, 5 $h^{-1}$Mpc,
10 $h^{-1}$Mpc, and 15 $h^{-1}$Mpc. The errors for all
the measures that have been made have been determined by dividing
the simulation box in eight equal  subsamples and by making eight different
measurements.

\vskip .5 cm
\leftline {\sl 5.1. Calculation of $S_p$ from N-body simulations}

The first test is the determination of the parameters
$S_3$ and $S_4$  compared to the theoretical predictions.
We calculated them from the counts of particles in the ensemble of $50^3$
spheres disposed on a grid. Thus, it corresponds to a spherical top-hat
filtering. As the number of particles is quite significant,
the shot noise effects are negligible and have been neglected to compute
the moments of the measured distributions.
 The resulting values of $S_3$ and $S_4$ are plotted in
Fig. 9 as a function of $\sigma$. For each filtering radius, we calculated
 the initial
effective index $n$ of the power spectrum to derive the expected value of
of $S_p(0)$ coefficients from formula (43). For the three filtering radii
  5 $h^{-1}$Mpc,
10 $h^{-1}$Mpc, and 15 $h^{-1}$Mpc we get correspondingly
$n\approx-1.3, -1.0, -0.7$.
Correspondingly, three values $S_3(0)\approx3.2, 2.9, 2.6$ and three
 values $S_4(0)\approx 17.7, 13.3, 10.6$  at  $\sigma=0$
plotted  in Fig. 9 without error bars  are these theoretical predictions.

 Each curve is related to a different filtering radius:
circles, squares and triangles correspond respectively to the smoothing with
 5 $h^{-1}$Mpc, 10 $h^{-1}$Mpc, and 15 $h^{-1}$Mpc. First point without
error bars on each curve is the theoretical prediction at  $\sigma=0$
we just discussed, which also can be interpreted as the initial time step
at redshift  $z \to \infty$.
Two other points on each curve correspond to two other time steps, at the
redshift  $z=0.6$  and at present $z=0$.
It can be seen that the theoretical prediction (43) -- how $S_p(0, n)$
depend on  the initial index $n$ -- is well reproduced by the extrapolation of
the numerical curves backward to $z \to \infty$, i.e. to $\sigma \to 0$.
Moreover, for a given smoothing radius, the $S_p$ parameters
do not exhibit any $\sigma$ dependence within the error bars.
This result makes the theoretical derivation of the $S_p(0)$
series particularly attractive.
It is also obvious that error bars are bigger for bigger filtering scale.
Other numerical results by Bouchet \& Hernquist (1992), and Lucchin \etal
(1993) indicate, however, that a variation with $\sigma$ may be
significative, especially in the nonlinear regime, and for low values of $n$.

\vskip .5 cm
\leftline {\sl 5.2. Calculation of PDF from N-body simulations}

The second numerical test is the construction of the density PDF
 from the N-body
simulation. We used the top-hat filtering of particle distribution at different
time steps $z=0.6$ and $z=0$ for different filtering radii  5 $h^{-1}$Mpc
and 15 $h^{-1}$Mpc. Top-hat smoothing allows us to construct
the PDF as count-in-cell statistics for the spherical geometry of cells.
Fig. 10 shows the Eulerian density PDFs at $z=0.6$ for two smoothing radii
 5 $h^{-1}$Mpc and 15 $h^{-1}$Mpc, which corresponds to $\sigma=0.92$ and
$\sigma=0.29$; as well as the PDFs at $z=0$ for the same smoothing but with
 $\sigma=1.52$ and
$\sigma=0.47$. This choice of parameters covers a broad range of non-linear
 stages,
$0.3 \lsim \sigma \lsim 1.5$. The error bars are the standard
deviation of the mean in eight
equal subsamples. We plot $P(\rhos)$ over wide range of density up to
$\rhos=10$.

We compare the numerical PDF with the two most advanced analytical predictions.
First, we plot the density PDF  from (27) derived in ZA without
smoothing, for four corresponding
values of $\sigma$.  Formula (27) approximates numerical PDF with top-hat
smoothing  very well up to $\sigma \lsim 0.5$ in the range $0 < \rhos \lsim 3$,
then starts to overestimate the high-density tail.
This is the range of validity
of the underlying assumptions of the Zel'dovich approximation
without final smoothing.
 For larger $\sigma$ the approximation (27) is slightly worsening,
and is out of applicability in the multiple
stream regime $\sigma > 1$. Our conclusion confirms of that of
Kofman \etal (1994), based on the Gaussian filtering.

We also plot the density PDF from  the exact perturbation theory in the single
 stream regime and with
the final smoothing. Assuming that  $S_p$ parameters are constant in
quasilinear regime,
 we substitute  the generating function $\varphi^S(y)$ from
 equation (42) into the
reconstruction formula (19), and calculate $P(\rhos)$ for
corresponding values of
$\sigma$ and $n$.
The results are presented in Fig. 10, and show a remarkable agreement with
the numerical  density PDF over the whole range of $\rhos$.
We extrapolate the theoretical PDF from the perturbation theory
for non-small $\sigma$
beyond the range of its validity. However,
the agreement with the numerical PDF is striking
  up to the maximal  used values $\sigma\approx 1.5$
and $\rhos \approx 10$, and so far there are no signs of deviation even for
higher $\sigma$! Plausible explanation of why formulae (19), (42) work so well
is that $S_p(\sigma, n)$ parameters depends very weakly on $\sigma$ up to
moderate $\sigma$,
at least for $n \gsim -2$.
 It has been explicitly checked for $S_3$ and $S_4$
but it should also be true for any of them otherwise the
low- and high-density  tails would not have been reproduced so accuratly.
Then it is not too surprising that
the density PDF (27) from ZA for which the $S_p$
parameters are not constant (see Fig. 6) does not fit well
 the low- and high-density tails for  $n\gsim -2$.
 For $n\lsim -2$, however, the
$S_p(\sigma)$ parameters may have a stronger dependence on $\sigma$,
and the ZA could then provide a more reliable PDF.
In any case it would be interesting to check  these theoretical
predictions against the PDF from N-body simulations for higher $\sigma$.

\vskip .5 cm
\leftline{\sl 5.3.  Fitting by the  log-normal distribution}

As it was noted a long time ago by Hubble (1934), the galaxy count
distribution in the plane cells on the sky might be well described
by the log-normal distribution. The log-normal distribution
fits the observed   galaxy PDF from 3D surveys as well
(Hamilton 1985, Bouchet \etal 1993, Kofman \etal 1994).
The log-normal density distribution reads as
$$P_{\rm log}(\rhos)\d\rhos=
{1\over\left(2\pi\sigma_{\rm l}^2\right)^{1/2}}
\exp\left[-{\left(\ln{\rhos}+\sigma_{\rm l}^2/2\right)^2\over
2\sigma_{\rm l}^2}\right]{\d\rhos\over \rhos},\ \ \sigma_{\rm l}^2=\ln
(1+\sigma^2).\eqno(44)$$
Kofman \etal (1994) found that the log-normal distribution is an excellent
approximation to the density PDF from N-body CDM simulation for moderate values
of  $\sigma$ in the used range $\rhos \le 5$.
In Fig. 10 we also compared the  density PDF from N-body CDM simulation with
 the log-normal distribution. We also found   a
striking agreement
between the log-normal PDF and that from N-body CDM simulation
 for the tested values of $0.3 < \sigma <1.5$  in the
tested range of $\rhos \le 10$!

Such a remarkable fitting inspires the thought that there might be a
strong dynamical reason to manifest the log-normal features of the
density PDF. For instance,  Coles \& Jones (1991) argued for the log-normal
mapping of the linear density field to describe its non-linear evolution.
Their log-normal model is universal for any spectral index $n$.
Unfortunately the log-normal mapping does not work (Coles \etal 1993).

Why does the log-normal density PDF work so well?

We argue  that the log-normal successful  fit can be
seen as a mere coincidence due to the shape of the
CDM power spectrum at moderate $\sigma$.
The log-normal PDF  is not a universal
form of the cosmic density PDF due to the non-linear dynamics, but is rather
a convenient fit for the particular region in the plane of
$(\sigma, n)$-parameters.
This region of  $(\sigma, n)$-parameters
includes the CDM model at moderate $\sigma$.
It explains why the log-normal PDF fits the results of N-body CDM simulations.
Consequently, the ``log-normalish'' features of the observed density PDf mean
that the realistic cosmological model corresponds to that
  $(\sigma, n)$-region,
i.e. close to the CDM model in this respect.

Let us consider the properties of the cumulants of the log-normal distribution.
The first two $S_p$ parameters of this distribution  as a function of
$\sigma$ are
$$\eqalignno{
S_3^{\rm log}(\sigma)&=3+\sigma^2,&(45a)\cr
S_4^{\rm log}(\sigma)&=16+15 \sigma^2+6 \sigma^4+\sigma^6,&(45b)\cr}$$
for an arbitrary $\sigma$.
For the sake of completness  note that  the log-normal distribution has
$S_p^{\rm log}(0)=p^{p-2}$, and its $\mG^{\rm log}(\tau)$-function is
simply $\exp(-\tau)$.

Parameters $S_3^{\rm log}(\sigma)$ and $S_4^{\rm log}(\sigma)$ are plotted
on Fig. 11 as dashed lines. We compare curves (45) with the values
of $S_3$ and $S_4$ from N-body CDM simulations for four values of
$\sigma= 0.29, 0.47, 0.92, 1.52$. The log-normal curves (45) are shown
to match the CDM parameters for moderate  values of $\sigma$  reasonably well.
The Edgeworth expansion (21) then helps to explain why the shape of the two
density PDFs -- from N-body CDM simulation and from (44) -- are in good
agreement if there is good agreement between $S_p$ parameters. Consequently,
we predict the agreement  worsening as far as $S_p$-s are departing
from each other. The accuracy of the log-normal  distribution is good
essentially due to the particular slope of the CDM power spectrum
at corresponding smoothing scales at moderate $\sigma$. A less steep
power spectrum, for instance,  would not have led to the same level of
agreement. The formula (42) with (19)  shows what  the expected dependence
of the density PDF with $n$ is (see Bernardeau 1994 for more details).
The log-normal distribution is expected to fail for the higher values of
$\sigma$ where the density PDFs from N-body simulations exhibit a power
law behavior (see for instance, Bouchet \& Hernquist 1993), which is
not the case for the form (44). The low- and high-density tails of the
log-normal distribution are very different from those given by the
reconstruction formula (19), as it is shown in the Table  below.

In practice, however,  for an observed power spectrum  $n\approx-1$ at
moderate $\sigma$, the log-normal distribution is a very effective and
simple fit for the density PDF in the mildly non-linear regime.
In accompanying paper (Kofman \& Bernardeau, 1994) we address
 the domain of validity of the log-normal fit more specifically.

\vskip .5 cm
\leftline{\bf 6. CONCLUSION}

\vskip .5 cm
\leftline{\sl 6.1.  Theoretical framework}

Let us recall the assumptions  that have been made  throughout
the paper to derive the various presented results.
As we noted in Section 3.1, the actual gravitational dynamics for the realistic
cosmological scenarios is quite complicated and includes the superposition
of hierarchical pancaking and clustering across the vast range of cosmological
scales. In the non-linear regime, the basic equations in terms of the
 continuous cosmic  fileds are multiple stream. However, at large scales,
ignoring small scale substructures, we expect that the
gravitational clustering is simpler, and might be approximated by the
single stream regime. Unfortunately, this transition
to the large--scale single stream description (i.e. the demonstration that,
indeed, the small scale details can be ignored for the large--scale dynamics)
has never been done, even in the linear regime.
The common (but formally unjustified yet) belief is that after the
large--scale filtering the perturbation theory, or truncated ZA can be applied.
It relates to a lesser extent to the N-body simulations with inevitable
truncation of the genuine power spectrum.
We are also  working within this assumption, basically because the results
we derive are in good agreement with the N-body simulations.
However, we clearly understand the nature of the approximation and
the need to justify it.

The usual practice is to filter the initial fluctuations in order to
model the large scale dynamics. Thus, our first assumption is that

1a) the initial fluctuations are smoothed
to ensure being in the single stream regime, the particular filtering scale $R$
is controlled by the rms density contrast, $\sigma=\sigma(R)$.

We used two  theoretical approaches (in particular, to derive
the density PDF). One  is to assume that

2a) the truncated Zel'dovich approximation can be used to describe the
large--scale dynamics.

The results which might be derived under these two assumptions are given
in \S 3. Note that the ZA can be used whatever the nature of the
initial conditions is, Gaussian or not.

The other approach is  the perturbation theory for the
cosmological equation of continuity, the Euler and the Poisson equations.
This method intrinsically relies on the
hypothesis of Gaussian initial conditions.
But in any case, even in the single stream approximation, the
equations are fully non-linear and it is difficult, if not impossible,
to find an exact solution. For instance, the general form of the $S_p(\sigma)$
parameters as  functions of $\sigma$ are beyond our current
skills of analytic calculations. We are then led to derive
them in their leading order in $\sigma$, that is

2b)  the cumulants are derived in the limit of small $\sigma$.

This calculation can be done rigorously  from the summarized
perturbation series. Note, that in the paper Bernardeau (1992)
the PDF was derived under the assumption 1a) and 2b), contrary to how
this paper is sometimes quoted.

So, within  the assumption 1a) and 2a) or 1a) and 2b)
it is possible to make interesting theoretical predictions, for instance,
get the dynamical derivations for the one-point density PDF and moments.
However, these calculations do not take into account
the final filtering that, in practice, cannot be avoided.
This is a crucial step since it alters the
statistical properties of the cosmic fields.
If the scale of the final filtering is larger than the scale of the initial
filtering, then the initial filtering is irrelevant, and we
can replace the assumption 1a) by the assumption that

1b)  this is the final smoothing only, which assures that the
large--scale dynamics can be accurately described by the single stream
approximation.

Even within the simplified dynamics given by the ZA
it is no longer possible to derive the density PDF under this last
assumption. The only known approach
that allows to do this
 comes from perturbative calculations,
since it makes it possible to derive the leading order of the cumulants
of the final density field. It is then a considerable
improvement compared to previous results. The assumptions 1b)
and 2b) then lead to a density PDF in very good agreement
with the numerical results.

%\page
\vskip .5 cm
\leftline{\sl 6.2.  Results}

We derived the density PDF, cumulants in forms of $S_p(\sigma)$-parameters and
their generating functions in case of 1D gravitational dynamics,
in the 3D Zel'dovich approximation with and without final smoothing,
in the perturbation theory extrapolated over mildly non-linear regime,
with and without final smoothing, for the log-normal distribution, and from
cosmological N-body CDM simulation. We summarize the quantitative
results in the Table.

We also present new  qualitative results
stemming from our study. As one can see from the Table,
the values
$S_p(0)$  are constants which characterize the  particular dynamical model.
For example, these values are different
for 1D gravitational instability, 3D Zel'dovich approximation, single stream
 3D gravitational instability, sea waves dynamics, etc.
The common trend is that $S_p$ is rapidly increasing with $p$, but at a
different rate. The $\sigma$-dependence of the
$S_p(\sigma)$ parameters  also characterizes the particular dynamics,
and as  we illustrated for different models,  might also be quite different.
 The smoothing effects introduces an extra
dependence on the power spectrum index $n$,
 $S_p(\sigma, n)$. We can expect $S_p(\sigma, n)$ from
different models can coincide in some regions of parameters $(\sigma, n)$.
It allows to construct simple fitting formula, as we found it for the
log-normal distribution.

$S_p(0)$-coefficients are related to the particular non-Gaussian statistics
which emerges from the non-linear dynamics. The usual practice of their direct
measurements  from observations is significantly affected by the final
sample volume. We, however, learned from the Edgeworth expansion that
$S_p(0)$ coefficients are related to the shape of the PDF peak. It gives us an
alternative method to evaluate the skewness and kurtosis by measuring the PDF
around its maximum, which is statistically more robust.
 This approach might be interesting in other contexts, such as
to constrain the skewness of the cosmological $\Delta T/T$ fluctuations,
or skewness and kurtosis of the cosmological velocity field and its divergence.

There is a deep connection between the generating function of
$S_p(0)$ parameters and  $\mG(\tau)$-function -- the generating function of
vertices, and the non-linear dynamics of the spherical overdense fluctuation
in the particular dynamical model. The solution of the last problem
in the particular dynamical model gives the  $\mG(\tau)$-function, which allows
to reconstruct the density PDF in the limit of small $\sigma$.
This is a remarkably simple and general prescription. From the Table
note a simple generalization of the $\mG(\tau)$ for
the models without final smoothing:
$\mG(\tau)={(1+\tau/\alpha)^{-\alpha}}-1.$
For the Zel'dovich approximation in the  space of $N$-dimensions,
$\alpha=N$. For the 3D perturbation theory $\alpha \approx 1.5.$
In the limit $N \gg 1$ we get $\mG(\tau)=\exp(-\tau) -1,$
which coincides with  that of the log-normal distribution.

We have derived the cosmological density PDF in two different approximations.
The one based on ZA gives  the density PDF
in the early non-linear regime, $\sigma \lsim .5$, and is expected
to improve for the models with $n\lsim -2$, for which pancaking is more
 pronounced.
This approach remains irreplaceable for models with non-Gaussian
initial density fluctuations. However,
we found that the best theoretical model for the density PDF
evolving from Gaussian distribution, in the case of  $n \gsim -2$,
 is based on the
perturbation theory when the final smoothing is included.
This PDF works remarkably well for significant range of $\sigma$ in
the mildly non-linear regime. Both approaches
 provide us with an efficient machinery
to deal with one-point statistics for a broad range of models.
In the Table  we summarize the properties of the various approximations
that have been presented and used in this paper.
It shows, in particular, that the low-density tail
is not affected by the reconstruction method.
 On the other hand, the shape of the high-density
tail is  dramatically modified. Both tails are also slightly modified,
 where the smoothing
effects are taken into account.

It was noted earlier and further confirmed that the log-normal distribution is
an excellent fit for the density PDF from CDM non-linear dynamics.
We found an explanation of this mystery, based on the properties of cumulants.
The log-normal distribution fits well in the particular range of
the parameter space $(n,\sigma)$ around $n \sim -1$, $\sigma \sim 1/2$,
and worsening outside of this region. By chance, the popular
CDM model at moderate $\sigma$ corresponds to this region.
Thus, some ``log-normal'' features of the observed density PDF would mean
that realistic cosmological model is close to that range of parameters.

\bigskip
\leftline{\bf Acknowledgements}

 We thank D.Bond and S.Shandarin for useful discussions,
 H.Couchman for providing
us the results of his N-body simulation,
 M. Longuet-Higgins for draw
attention to the Edgeworth series for slightly non-Gaussian statistics.
L.K. thanks the support from the CIAR cosmological program, F.B. thanks
the University of Hawaii for hospitality.

\vfill\eject
\leftline{\bf APPENDIX A:  Statistical   technics   }

In this appendix, we give technical definitions
of interest for statistical studies, such as the moments of
the density distribution function, its cumulants, the generating functions
of the moments and of the cumulants. Relationship between those
quantities are also given.
-1z
The moments $\mg\delta^p\md$ of the distribution function, $P(\rho)$,
 are given by the integrals,
$$\mg\delta^p\md=\int_0^{\infty}\d\varrho\ P(\varrho) (\varrho-1)^p.\eqno(A1)$$
The cumulants can then be obtained from the moments.
The $p^{th}$ cumulants, $\mg\dta^p\md_c$, is
defined recursively from the $p^{th}$ moment
following the rules,
$$\eqalign{
\mg\dta\md_c&=0,\cr
\mg\dta^2\md_c&=\mg\dta^2\md\equiv \sigma^2,\cr
\mg\dta^3\md_c&=\mg\dta^3\md,\cr
\mg\dta^4\md_c&=\mg\dta^4\md-3\mg\dta^2\md_c^2,\cr
\mg\dta^5\md_c&=\mg\dta^5\md-10\mg\dta^2\md_c\mg\dta^3\md_c,\cr
\dots\cr}\eqno(A2)$$
In general, to obtain the $p^{th}$ cumulant, one should consider all the
decompositions of a set of $p$ points in its subsets (but the one being only
the set itself); multiply, for each decomposition, the cumulants
corresponding to each subset and subtract the results of all these products
 obtained in that way from the $p^{th}$ moment.

Each cumulant, in a given order, actually contains
a piece of information about the shape of the PDF that cannot
be derived from the lower order cumulants. For instance, the second cumulant
gives the width of the distribution, the third cumulant measures its
asymmetry, the fourth -- its flatness, etc. The Gaussian distribution
is characterized by only one non-zero cumulant (whereas all even
moments are nonzero): the second one,
which gives its variance. However, in general, a distribution function
will be characterized by the whole  series of cumulants.
Therefore, to treate more easily  the series of the cumulants,
we are led to define other mathematical tools of great practical interest.
There are the generating functions of the moments and  the
cumulants. The generating function of the moments
is defined by,
$$\mM(\mu)=1+\sum_{p=1}^{\infty}\mg\dta_p\md {\mu^p\over p!}.\eqno(A3)$$
The generating function of the cumulants, $\mC(\lambda)$,
is defined in a similar way,
$$\mC(\mu)=\sum_{p=2}^{\infty}\mg\dta_p\md_c {\mu^p\over p!}.\eqno(A4)$$

One fundamental result of the statistics is that $\mM(\mu)$ and $\mC(\mu)$
are connected in a simple way. Indeed, we have
$$\mM(\mu)=\exp[\mC(\mu)].\eqno(A5)$$
We omit here the rigorous demonstration of this property
(see, e.g. Balian \& Schaeffer 1989).
By expanding $\exp[\mC(\mu)]$ with respect to $\mu$,
one can easily verify that the first few moments are  given correctly.

Using this property, it is then possible to relate the generating function
of the moments, or of the cumulants, to the shape
of the PDF. Indeed, we have
$$\eqalign{
\mM(\mu)\exp\left[\mC(\mu)\right]
&=\sum_{p=0}^{\infty}\int_0^{\infty}\d\varrho\ P(\varrho)
{\left[(\varrho-1)\mu\right]^p\over p!}\cr
&=\int_0^{\infty}\d\varrho\ P(\varrho)\exp[(\varrho-1)\mu].\cr}\eqno(A6)$$

\vfill\eject
\leftline{\bf APPENDIX B: Derivation of the Edgeworth expansion}

In this appendix we present the derivation of the Edgeworth
expansion. It is based on the reconstruction (19)
for the density PDF. The generating function $\varphi(y)$
can be expanded with respect to $y$ (eq. [20]). We then have to expand
the non-Gaussian part of the exponent in (19),
$$\eqalign{
\exp&\left[-{\varphi(y)\over\sigma^2}\right]
\approx \exp\left[{-y+y^2\over2\sigma^2}\right]\cr
\times&\left[1-{S_3\over 3!}{y^3\over \sigma^2}
+{S_4\over 4!}{y^4\over\sigma^2}+{S_3^2\over 2 (3!)^2}{y^6\over\sigma^4}
-{S_5\over 5!}{y^5\over\sigma^2}-{S_3\ S_4\over 3!\,4!}{y^7\over\sigma^4}
-{S_3^3\over (3!)^4}{y^9\over\sigma^6}+
\dots \right].\cr}\eqno(B1)$$
This expansion should be made with respect to both $y$ and $\sigma$,
assuming they are of the same order.
The resulting value of the density PDF requires the determination
of integrals of the form
$$I_n=\int_{-\i\infty}^{+\i\infty}{\d y\over 2\pi\i\sigma^2}
\exp\left[{y^2\over2\sigma^2}+{(\varrho-1) y\over \sigma^2}\right]
y^n,\eqno(B2)$$
(for $n\ge 3$) that gives
$$I_n={(-1)^n \over (2\pi\sigma^2)^{1/2}}
\exp\left(-\nu^2/2\right)
\ \sigma^n\ H_n\left(\nu\right),\ \ \ \ \ \nu=\dta/\sigma\eqno(B3)$$
where $H_n(\nu)$ are the Hermite polynoms,
$$\eqalign{
H_n(\nu)&\equiv(-1)^n\exp(\nu^2/2) {\d^n\over\d\nu^n}\exp(-\nu^2/2)\cr
&=\nu^n-{n(n-1)\over 1!}\ {\nu^{n-2}\over 2}
+{n(n-1)(n-2)(n-3)\over 2!}\ {\nu^{n-4}\over 2^2}-\dots\cr}\eqno(B4)$$
The resulting form of the density PDF is
$$
\eqalign{
P(\delta)\d\dta=&{1 \over (2\pi\sigma^2)^{1/2}}
\exp
\left( -\nu^2/2\right)
\biggl[1 + \sigma {S_3 \over 6} H_3\left(\nu\right)
+\sigma^2\biggl( {S_4 \over 24} H_4\left(\nu\right)
+{S_3^2\over72}H_6\left(\nu\right)
 \biggr)\biggr.\cr
&\biggl.+\sigma^3\biggl({S_5\over 120} H_5(\nu)+
{S_4 S_3\over 144} H_7(\nu)+ {S_3^3\over 1296} H_9(\nu)\biggr)
+ ... \biggr] \d\dta.\cr}\eqno(B5)$$

\vfill\eject
\leftline{\bf APPENDIX C: $S_p(0)$ from the Perturbation Theory }

In this appendix we give a brief sketch of the derivation of some technical
details used Sec. 4.1.
Cosmological
 gravitational instability of the perfect fluid without pressure
in the single stream regime is described by the continuity equation
$$
{\partial \rho \over \partial t} +3H\rho +
 {1\over a}\nabla_{\rm x}(\rho \vv)=0,   \eqno(C1a)$$
the Euler equation
$$
{\partial \vv \over \partial t} +
 {1 \over a}(\vv \nabla_{\rm x} )\vv +H\vv = -{1 \over a}
\nabla_{\rm x} \phi, \eqno(C1b)$$
and the Poisson equation
$$
\nabla_{\rm x}^2\phi = 4\pi G a^2 (\rho - \bar \rho). \eqno(C1c)
$$
It is convenient to use $a(t)$ as a new time variable, and the
velocity potential $\Phi$ for the potential flow under consideration,
see \S 3.1 for definition. Let us consider the case
of the Einstein-de~Sitter Universe.
Then three equations (C1) can be rewritten in form of two equations:
$$
a{\partial \over \partial a}\dta(\vx,a)+(1+\dta(\vx,a))\Delta\Phi(\vx,a)+
\grad\dta(\vx,a).\grad\Phi=0\eqno(C2a)
$$
and
$$\eqalign{
a{\partial \over \partial a}\Delta\Phi(\vx,a)+{1\over 2}\Delta\Phi(\vx,a)+
\grad\Phi(\vx,a).\grad(\Delta\Phi(\vx,a))&\cr
+\sum_{\alpha,\beta=1}^{3}\Phi_{,\alpha\beta}(\vx,a)\Phi_{,\alpha\beta}(\vx,a)
+{3\over2}\dta(\vx,a)&=0,\cr}\eqno(C2b)
$$
with $\Phi_{,\alpha\beta}(\vx,a)=\left({1/ aH}\right)^2
{\partial^2\Phi(\vx,a)/ \partial \vx_{\alpha}
\partial \vx_{\beta}}$ where $\vx_{\alpha}$ is the component
${\alpha}$ of $\vx$.

Let us  seek the solution of the equations (C2) in the form of
perturbation series,
e.g. $\delta =\sum_{p=1}^{\infty}\delta^{(p)}$,
$\nabla^2 \Phi =\sum_{p=1}^{\infty}(\nabla^2\Phi)^{(p)}$, etc.
Then we obtain the hierarchy of equations for values $\delta^{(p)}$,
$(\nabla^2\Phi)^{(p)}$ etc., in each order $p$ of the perturbation series.
Then, at the lowest order of $\sigma$,
the cumulant $\mg\dta^p\md_c$ is given by a term in the form
$$\mg\dta^p\md_c=\sum_{{\rm combinations,}\ p(i)}
\mg\prod_{i=1}^p\dta^{(p[i])}\md_c\eqno(C3a)$$
where the sum is taken over all the possible combinations $p(i)$, $i=1\dots p$,
for which
$$p(i)\ge1,\ \ \ \sum_{i=1}^p p(i)=2p-2.\eqno(C3b)$$
This particular result is correct due to the hypothesis of the Gaussian initial
conditions. The terms in (C3b) are all products of
the rms density fluctuation at the power $2p-2$ by some
combination of the vertices,
$$\nu_p \equiv \mg\delta^{(p)} (\delta^{(1)})^n\md_c/\sigma^{2p}.\eqno(C4)$$
(and respectively $\mu_p$ for $\nabla^2 \Phi$, etc.)
The parameters $S_p$ (eq. [7]) for $\sigma=0$
are then given by a combination of the
vertices, $\nu_p$, that technically corresponds to a tree summation.
The relationship between the $S_p$ parameters and the vertices can be
written in a closed form at the level of their corresponding generating
functions.
Let us define the function $\mG(\tau)$ by
$$
\mG(\tau)=\sum_{p=1}^{\infty} {\nu_p \over p!}(-\tau)^p. \eqno(C5)
$$
(a similar one can be defined for other vertices).
Then we have
$$\varphi(y,0)=y+y\mG(\tau)+{\tau^2\over 2},\ \
\tau=-y{\d\over\d\tau}\mG(\tau),\eqno(C6a)$$
where $\varphi(y,\sigma)$ is defined in (20).

The problem then is to derive the function
$\mG(\tau)$ from the equations (C2).
F.B. 1992 derived useful properties of the generating functions of
vertices.
Multiplying equations (C2a)  by $(\delta^{(1)})^p$ and averaging
the result, and repeating similar operation with eq. (C2b),
 and then using properties
of the generating functions of vertices,
 one can finally get
 the single equation for $\mG(\tau)$:
$$\eqalign{
&-(1+\mG)\tau^2{d^2 \over d\tau^2}\mG+ {4\over 3}
\left(\tau{d \over d\tau}\mG\right)^2-
{3\over 2}(1+\mG)\tau{d \over d\tau}\mG+{3\over 2}
\mG (1+\mG)^2=0.\cr
&\ \ \ \ \
\hbox{with}\ \ \mG(\tau)\sim -\tau\ \ \ \hbox{when}\ \ \tau\to 0.\cr}
\eqno(C7)$$
It is remarkable, that this equation does not contain any space derivatives,
and has {\it exactly} the same form as the equation of the spherically
symmetric collapse of the ``overdense'' $\mG(\tau)$ with a ``scalar factor''
$\tau$.
The  analytical solution of the equation (C7) is well known.
When $\tau<0$
$$
\mG={9\over2}{(\theta-\sin\theta)^2 \over (1-\cos\theta)^3}-1,\ \ \
\tau=-{3\over5}\left({3\over4}(\theta-\sin \theta)\right)^{2/3}
\eqno(C8a)$$
and when $\tau>0$
$$
\mG={9\over2}{(\sinh\theta-\theta)^2 \over (\cosh\theta-1)^3}-1,\ \ \
\tau={3\over5}\left({3\over4}(\sinh\theta-\theta)\right)^{2/3}.
\eqno(C8b)$$

It turns out that the form (38) for
$\mG(\tau) \approx {1/( 1+\tau/1.5)^{1.5}}-1$ is a very
good fit to the exact solution (C8).

In general case of arbitrary background cosmology
the function $\mG(\tau)$ depends on the
cosmological parameters. Fortunately, this dependence is  weak,
and  the form (38) can be accurately used in
general case.

\vfill\eject
\leftline{\bf REFERENCES}

\apjref Alimi, J.M., Blanchard, A. \& Schaeffer, R. 1990; ApJ; 349; L5;
\apjref Balian, R. \& Schaeffer, R. 1989; A\&A; 220; 1;
\apjref Bernardeau, F. 1992; ApJ; 392; 1;
\prepref Bernardeau, F. 1993; CITA 93/44 preprint;
\prepref Bernardeau, F. 1994; CITA 94/7 preprint;
\apjref Bernardeau, F. \& Schaeffer, R. 1992; A\&A; 255; 1;
\prepref Bernardeau, F., Singh, T.P., Banerjee, B. \& Chitre, S.M.; 1994,
MNRAS in press;
\bookref  Bond, J.R., \& Couchman, H. 1988; in  Proc. 2-th
Canadian Conf. on GR and Rel. Astrophysics; eds. Coley A.
\& Dyer C.; World Scientific, p. 385;
\apjref Bouchet, F. \& Hernquist, L. 1992; ApJ; 400; 25;
\apjref Bouchet, F., Juszkiewicz, R., Colombi, S., \& Pellat, R. 1992; ApJ;
394; L5;
\apjref Bouchet, F., Schaeffer, R. \& Davis, M. 1991; ApJ; 383; 19;
\apjref Bouchet, F., Strauss, M.A., Davis, M., Fisher, K.B., Yahil,
A. \& Huchra, J.P. 1993; ApJ; 417; 36;
\prepref Catelan, P. \& Moscardini, L. 1993; preprint;
\apjref Coles, P. \& Jones, B. 1991; MNRAS; 248; 1;
\apjref Coles, P., Melott, A. \& Shandarin, S. 1993; MNRAS; 260; 765;
\prepref  Colombi, S.,  Bouchet, F., \& Schaeffer, R. 1993; A\&A submitted;
\apjref Couchman, H.M.P., 1991; ApJ; 368; L23;
\apjref Davis, M. \& Peebles, P.J.E. 1977; ApJS; 34; 425;
\apjref Doroshkevich, A.G., 1970; Astrofizica; 6; 581;
\apjref Fry, J. 1984; ApJ; 279; 499;
\apjref Fry, J.N. \& Gazta\~naga, E. 1992; ApJ; 413; 447;
\apjref Gazta\~naga, E. \& Yokoyama, J. 1993; ApJ; 403; 450;
\apjref Goroff, M.H., Grinstein, B., Rey, S.-J. \& Wise, M.B. 1986;
ApJ; 311; 6;
\apjref Grinstein, B. \& Wise, M.B. 1987; ApJ; 320; 448;
\bookref Gurbatov, S., Malahov, A. and Saichev, A. 1991; Nonlinear Random Waves
and Turbulence in Nondispersive Media; Manchester University Press;
\apjref Hamilton 1985; ApJ; 292; L35;
\apjref Hubble, 1934; ApJ; 75; 8;
\apjref Juszkiewicz, R. \& Bouchet, F. \& Colombi, S. 1993; ApJ; 419; L9;
\prepref Juszkiewicz, R., Weinberg, D. H., Amsterdamski, P., Chodorowski, M.
\& Bouchet, F. 1993; IASSNS-AST 93/50 preprint;
\bookref Kofman, L. 1991; in Primordial Nucleosynthesis \& Evolution of Early
Universe; eds. Sato, K. \& Audouze, J., (Dordrecht: Kluwer);
\prepref Kofman, L.A. 1993; astro-ph/9311027;
\prepref Kofman, L.A. 1994; in preparation;
\prepref Kofman, L. \& Bernardeau, F. 1994; in preparation;
\apjref Kofman, L.A., Melott, A., Pogosyan, D.Yu. \& Shandarin, S.F. 1992; ApJ;
393; 437;
\apjref Kofman, L., Bertschinger, E., Gelb, M.J., Nusser, A., Dekel, A. 1994;
ApJ; 420; 44;
\apjref Longuet-Higgins, M., 1963; J. Fluid Mech; 17; 459;
\apjref Lucchin, F., Mataresse, S., Melott, A.L. \&
Moscardini, L. 1994; ApJ; 422; 430;
\prepref Munshi, D. \& Starobinsky, A. 1993; preprint;
\apjref Padmanabhan, T., \& Subramanian, K. 1993; ApJ; 410; 482;
\bookref Peebles, P.J.E. 1980;  The Large--Scale Structure of the Universe;
Princeton University Press, Princeton, N.J., USA;
\bookref Peebles, P.J.E. 1993;  Principles of Physical Cosmology;
Princeton University Press, Princeton, N.J., USA;
\apjref Saslaw, W.C. \& Hamilton, A.J.S. 1984; ApJ; 350; 492;
\apjref  Shandarin, S.F. and Zel'dovich, Ya.B., 1989;  Rev.Mod.Phys.; 61; 185;
\bookref Shandarin, S.F. 1992; Proceedings of the IUCAA conference, Dec. 1992;
Pune, India;
\apjref Weinberg, D. \& Cole, S. 1992; MNRAS; 259; 652;
\apjref Zel'dovich, Ya.B. 1970; A\&A; 5; 84;
\apjref Zeldovich, Ya.B., Shandarin, S.F., 1982; Sov.Astron.Lett.; 8; 139;

\page
\leftline{\bf Table and Figure captions}
{\noindent

{\bf Table 1.}
Properties of the density PDF obtained with
the various approximations described in the text. The first two columns
give the values of the $S_3$ and $S_4$ parameters. In general case
 they depend on $\sigma$. A couple of first terms
of the $\sigma$-expansion are given in cases they are known.
 The third column gives the shape
 of the function
$\mG(\tau)$ which is  used to derive the function $\varphi(y)$,
involved in the reconstruction formula (19).
The last two columns give the high- and low-density asymptotas of the
resulting distributions.
\vskip .5 cm

{\bf Fig 1.}
Shape of the density PDF for 1D dynamics and for $\sigma=0.5$ (eq. [4])
(solid line). The dotted, dashed, long dashed and dotted dashed lines
show the shape of the density PDF defined in Eq. (6) for
respectively $\varrho_c$=5, 7.5, 10, 12.5.
\vskip .5 cm

{\bf Fig 2.}
coefficients after regularization for the 1D dynamics.
The filled circles correspond to the theoretical limits for
$\sigma\to 0$ (Eqs. [15]),
and the thick solid lines correspond to the theoretical curves (Eqs. [16]).
The dotted, dashed, long dashed and dotted dashed lines are
for a sharp cutoff (Eq. [5], left panel) or
an exponential  cutoff (Eq. [6], right panel)
for respectively $\varrho_c=$5, 7.5, 10, 12.5.
\vskip .5 cm

{\bf Fig 3.}
The shape of the density PDF as obtained from eq. (4), thin solid lines,
 and by
the reconstruction method [eq.(19)], thick solid lines,
for $\sigma=0.3$ (left panel) and $\sigma=0.5$ (right panel).
\vskip .5 cm

{\bf Fig 4.}
The Edgeworth expansion (Eq. [22]) compared to the form (4)
for $\sigma=0.1$ (left panel) and $\sigma=0.3$ (right panel).
The dashed line corresponds to the case  when the skewness only
is taken into account (up to $\sigma$ correction), the long dashed lines
when both the skewness and the kurtosis are taken into account,
up to $\sigma^2$ corrections,
and the dotted line when the expansion is made up to the $\sigma^3$
order (Eq. [22]).
\vskip .5 cm

{\bf Fig 5.}
Shape of the density PDF for $\sigma=0.3$ (left panel) and
$\sigma=0.5$ (right panel) for various methods based
on the Zel'dovich approximation.
The solid line is the shape obtained by the
Zel'dovich approximation, (Eq. [27]); the other curves have
been obtained by the reconstruction formula (19)
 by assuming  that the ratios $\mg\dta^p\md_c/\mg\dta^2\md^{p-1} =S_p^Z$
equal their low-$\sigma$ limit in the ZA without taking into
account the smoothing effects (dashed lines), and taking into
account the smoothing effects for $n=-1$ (long dashed lines).
\vskip .5 cm

{\bf Fig 6.}
parameters after regularization for the 3D dynamics.
The dotted, dashed, long dashed and dotted dashed lines are
for the sharp cutoff (left panel) or the exponential  cutoff (right panel)
for respectively $\rho_c=$5, 7.5, 10, 12.5 in the regularized expression (27).
The filled circles correspond to their theoretical limit at
small $\sigma$ and the squares to their theoretical
limits when the filtering effects are taken into account,
for $n=-1$.
\vskip .5 cm

{\bf Fig 7.}
Shape of the density PDF for $\sigma=0.3$ (left panel) and
$\sigma=0.5$ (right panel) from the reconstruction
formula (19) using the value of the cumulants from (38) for the exact dynamics
in the $\sigma\to 0$ limit.
The solid line is the density
PDF obtained by assuming that the values $S_p$
equal their low $\sigma$ limit without taking
into account the smoothing effects. The dashed
line is obtained when the smoothing effects are taken into account
assuming that the initial power spectrum is $P(k)\propto k^{-1}$.
The long dashed line is the  log-normal distribution.
\vskip .5 cm

{\bf Fig 8.}
The Edgeworth expansion (Eq. [21]) compared to the density PDF
obtained with (42) and the reconstruction formula (19) for $n=-1$,
$\sigma=0.3$ (left panel) and $\sigma=0.5$ (right panel).
The dashed line corresponds to the case  where the skewness only
is taken into account (up to $\sigma$ correction), and the long dashed lines
to the case where  both the skewness and the kurtosis are taken into account,
up to $\sigma^2$ corrections. The long dashed line is the lognormal
distribution.
\vskip .5 cm

{\bf Fig 9.}
The coefficients $S_3(\sigma)$ and $S_4(\sigma)$ as  functions
of $\sigma$ in a CDM numerical simulation. The  circles, squares
and triangles correspond respectively to the smoothing radii of
15, 10, 5 $h^{-1}$Mpc.
Three points at each curve corresponds to three different time steps,
at $z=\infty, 0.6, 0$.
 The values at  $\sigma=0$ (or $z \to \infty$) are the
theoretical predictions from (43) taking into account the filtering effects.
\vskip .5 cm

{\bf Fig 10.}
The density PDF for CDM initial conditions. The
points are measured in a numerical simulation at two different
time steps corresponding to $a/a_0=0.6$ (upper panel) and $a/a_0=1$
(lower panel) and at two different smoothing radii $R_0=5h^{-1}$Mpc
and $R_0=15h^{-1}$Mpc. The rms density fluctuation are then respectively
$\sigma=0.92$ and $\sigma=0.29$ in the upper panel and $\sigma=1.52$
and $\sigma=0.47$ in the lower panel. The error bars have been obtained by
dividing the sample into eight subsamples. The solid line is the prediction
 given
by (19) when the smoothing effects are taken into account (with Eq. [42]).
The dashed line is the prediction (27) from the ZA and
the long dashed line is the lognormal distribution (44).
\vskip .5 cm

{\bf Fig 11.}
The measured values of the $S_3(\sigma)$ and $S_4(\sigma)$ parameters in
 the CDM
simulation as  functions of $\sigma$.
The solid lines correspond to the values of these coefficients
from formula (43) for
 $n=-1$.
 The thick dashed lines correspond to the log-normal distribution,
eq. (45). The  $S_3$ and $S_4$ values in CDM and log-normal models overlap
 for moderate $\sigma \sim 0.5$ only.
\vskip .5 cm
}

\bye